\documentclass[]{IEEEtran}

\usepackage[noadjust]{cite}
\usepackage[colorlinks=false,linkcolor=black,urlcolor=black,bookmarksopen=true, hidelinks]{hyperref}
\usepackage{algorithm} 
\usepackage{amssymb}
\usepackage{algorithmic}  
\usepackage[algo2e]{algorithm2e} 
\SetAlFnt{\small}
\SetAlCapFnt{\large}
\SetAlCapNameFnt{\large}
\usepackage{algorithmic}
\algsetup{linenosize=\tiny}
\usepackage{multirow}
\usepackage{graphicx}
\usepackage{indentfirst}
\usepackage{xspace}
\usepackage{amsmath}
\usepackage{blkarray, bigstrut}
\usepackage{tabstackengine}
\usepackage{xcolor,soul}
\usepackage{adjustbox}
\usepackage{bm} 
\usepackage{color} 
\usepackage{array}
\usepackage{balance}
\usepackage{multirow}
\usepackage{marginnote} 
\usepackage[norndcorners,customcolors,nofill]{hf-tikz}
\usepackage{amsmath,times,tabstackengine}
\usepackage{booktabs}
\usepackage{multirow}
\usepackage{siunitx}
\SetKwComment{Comment}{/* }{ */}
\usepackage{comment}

\DeclareMathAlphabet\bmcal{OMS}{cmsy}{b}{n}

\usepackage{cite}
\usepackage{amsmath,amssymb,amsfonts}
\usepackage{authblk}
\usepackage{graphicx}
\usepackage{blkarray, bigstrut}
\usepackage{xspace}
\usepackage{xparse}
\usepackage{tabstackengine}
\usepackage{xcolor,soul}
\usepackage[flushleft]{threeparttable}
\newcolumntype{R}[1]{>{\raggedleft\let\newline\\\arraybackslash\hspace{0pt}}m{#1}}

\ifCLASSINFOpdf
\else
\fi
\begin{document}
%


\title{Time-Distributed Feature Learning for\\ Internet of Things Network Traffic Classification}



\author{Yoga Suhas Kuruba Manjunath, \textit{Graduate student Member, IEEE}, Sihao Zhao, \textit{Senior Member, IEEE}, \\
		Xiao-Ping~Zhang, \textit{Fellow, IEEE}, Lian Zhao, \textit{Fellow, IEEE}
\thanks{This work was supported in part by the Natural Sciences and Engineering Research Council of Canada (NSERC), Grant No. RGPIN-2020-04661. \textit{(Corresponding author: Xiao-Ping Zhang)}}
\thanks{
Y. Manjunath and L. Zhao are with the Department of Electrical, Computer and Biomedical Engineering, Toronto Metropolitan University, Toronto, ON M5B 2K3, Canada. (e-mail: yoga.kuruba@torontomu.ca, l5zhao@torontomu.ca).\\
X.-P. Zhang is with Shenzhen Key Laboratory of Ubiquitous Data Enabling, Tsinghua Shenzhen International Graduate School, Tsinghua University, and with the Department of Electrical, Computer and Biomedical Engineering, Toronto Metropolitan University, Toronto, ON M5B 2K3, Canada. (e-mail: xpzhang@ieee.org).\\ 
S. Zhao is with NovAtel, Autonomy Positioning division of Hexagon, Calgary, AB T3K 2L5, Canada (e-mail: zsh01@tsinghua.org.cn).

}
}
\maketitle

\begin{abstract}

Deep learning-based network traffic classification (NTC) techniques, including conventional and class-of-service (CoS) classifiers, are a popular tool that aids in the quality of service (QoS) and radio resource management for the Internet of Things (IoT) network. Holistic temporal features consist of inter-, intra-, and pseudo-temporal features within packets, between packets, and among flows, providing the maximum information on network services without depending on defined classes in a problem. Conventional spatio-temporal features in the current solutions extract only space and time information between packets and flows, ignoring the information within packets and flow for IoT traffic. Therefore, we propose a new, efficient, holistic feature extraction method for deep-learning-based NTC using time-distributed feature learning to maximize the accuracy of the NTC. We apply a time-distributed wrapper on deep-learning layers to help extract pseudo-temporal features and spatio-temporal features. Pseudo-temporal features are mathematically complex to explain since, in deep learning, a black box extracts them. However, the features are temporal because of the time-distributed wrapper; therefore, we call them pseudo-temporal features. Since our method is efficient in learning holistic-temporal features, we can extend our method to both conventional and CoS NTC. Our solution proves that pseudo-temporal and spatial-temporal features can significantly improve the robustness and performance of any NTC. We analyze the solution theoretically and experimentally on different real-world datasets. The experimental results show that the holistic-temporal time-distributed feature learning method, on average, is 13.5\% more accurate than the state-of-the-art conventional and CoS classifiers.

\end{abstract}


%
\IEEEpeerreviewmaketitle

\section{Introduction}


The Internet of Things (IoT) is an architecture with connected devices that can exchange data among themselves and with the Internet \cite{2015TowardsAD}. The technology is prevalent with many IoT nodes and abundant services \cite{Norton2019}. Therefore, naturally, the network traffic in IoT networks is growing dramatically. The rise in Internet traffic with an annual rate of 55\% by 2030 \cite{itc}, according to the International Telecommunication Union (ITU), signifies the importance of quality of service (QoS), radio resource management (RRM), and network resource scheduling in IoT networks.

Tuning the network for better QoS is the Internet service providers' (ISPs) job; therefore, they use network traffic management (NTM) as a fundamental tool to understand the behaviour of IoT traffic \cite{robitza2017challenges}. NTM is essential for RRM and scheduling processes as well. NTM comprises different components, including network traffic provision, classification, policy, resource management, and other functionalities. In this work, we focus only on the classification aspect of the NTM. Network traffic classification (NTC) is one of the main components of NTM that helps to segregate the traffic for further analysis \cite{pacheco2018towards}. NTC has different types, such as conventional, called multi-class or fine-grain, and class-of-service (CoS), or heap, NTCs. ISPs use both classifiers for different purposes in IoT NTM. Therefore, conventional and CoS classifiers are imperative tools for tuning IoT networks for better QoS.

The conventional NTC helps identify specific application-level traffic such as YouTube, Facebook, Google, domain name system (DNS), dynamic host configuration protocol (DHCP) and many others. In layperson's terms, a conventional traffic classifier is called fine-grain classification. Class-of-services (CoS) traffic classifiers classify traffic into a group of similar services such as P2P, video, chat and many more. For example, Facebook chat, Google chat (Hangout) and Skype chat are identified as chat by the CoS classifier. A CoS classifier is called a heap classifier in layperson's terms. Given the dynamic nature of IoT network traffic transactions, conventional and CoS NTCs are helpful tools for ISPs. Mainly, conventional NTC plays a vital role in service level agreement (SLA) verification and the lawful interception (LI) of illegal or critical traffic based on geography restrictions \cite{IETF3924}. The CoS classification is necessary for faster resource scheduling for a better QoS \cite{mehta2017survey} in IoT networks because knowing a type of service is sufficient instead of identifying the exact traffic service to tune QoS. The CoS classifier is also used in UE Route Selection Policy (URSP) \cite{usrp}. Therefore, having a unique classifier is beneficial for ISPs to realize hassle-free and quick network traffic management. 

\color{black}

Most of the current works \cite{9590566,liu2021toward, sivanathan2020detecting,yao2019capsule} deal with either conventional traffic classification or CoS classification for IP networks. Even though the solutions must hold good for IoT traffic since it is a subset of IP traffic, IoT traffic majorly differs with the characteristics of the application layer protocol and node's behaviour. Spatio-temporal features used in the current solutions are limited to representing IP traffic but need to be more adequate for IoT traffic. Also, there is no unique solution to handle both types of classification for IoT traffic. Therefore, it takes time and effort for ISPs to employ different NTCs based on their needs. The problem emphasizes the requirement of a unique classifier for ISPs to handle the IoT networks.

The recent works \cite{aceto2019mobile,9626148,9707490,9841019,10082962} have shown the success of deep learning in the NTC. The well-known deep learning architectures used in the NTC are the convolutional neural networks (CNNs) and long short-term memory (LSTM) recurrent neural networks. The hybrid structures such as CNN-LSTM are also explored in the NTC \cite{lopez2017network,d2021network,manjunath2021time}. However, the CNN, LSTM, and CNN-LSTM variants extract spatial, temporal, and spatio-temporal features for NTC. More than spatial-temporal features are needed to capture the dynamic nature of IoT traffic. Therefore, we need holistic-temporal features to characterize the IoT network traffic \cite{manjunath2021time}. 

Regular traffic flows primarily exhibit intra- and inter-temporal characteristics \cite{tian2020chaotic,8409450}, nothing but spatio-temporal information, which explains the performance of current deep learning solutions for IP traffic and failure for IoT traffic because of a lack of holistic-temporal features. Also, using spatio-temporal features binds deep learning models to the problem defined; therefore, current deep learning methods are limited to either conventional or CoS problems. Holistic-temporal features help represent IoT network traffic efficiently. IoT network traffic exhibits stationarity and non-stationarity as a session progresses \cite{ntlangu2017modelling}; therefore, we need more information than just spatio-temporal information to capture more comprehensive features of the traffic. The holistic-temporal features contain - i) intra-temporal features, which are the temporal relationships within the network flow; ii) inter-temporal features, which are the temporal information among flows; and iii) pseudo-temporal features, which are the relation between intra and inter-temporal features. We call it pseudo-temporal features since it is of a temporal nature that is mathematically complex to explain since it is extracted by deep learning - a black box. The intra-temporal relation can be exhibited by sequence number acknowledgements, timestamps, time to live (TTL), etc. The behavioural study of traffic services also unveils the inter-temporal relation among flows \cite{9979671}. Pseudo-temporal features capture the interrelation among intra- and inter-temporal features. Therefore, using holistic-temporal features can solve both conventional and CoS problems without depending on the raw features used in the problem.

In summary, we identify two main research gaps in the current solutions: i) lack of single solution for multi-class and CoS traffic classifications for IoT traffic, and ii) need for work in extracting holistic-temporal features to maximize the classifiers performance irrespective of the initial features. The other drawbacks are the robustness and generalization of current classifiers since those are tested on limited and small datasets. The main goal of our work is to address the above-identified issues.


In this paper, we propose a novel solution that uses a time-distribution wrapper over CNN-LSTM for holistic-temporal feature extraction and thereby achieve a standard feature engineering method for both types of classification. Also, to enable time-distribution feature learning, we propose a novel method of traffic data representation. Our solution employs a CNN to extract the intra-temporal (or spatial) features within a network flow statistical data. We utilize a long-term memory recurrent neural network (RNN) to extract the inter-temporal features among the network flows statistical data. Finally, a time-distributed wrapper is used over a feed-forward connected neural network to enable pseudo-temporal feature learning. Time-distributed feature learning is realized by representing the traffic data as a greyscale video stream. We prove the efficiency of the proposed work using theoretical and experimental analysis. 

The proposed method is evaluated using diverse real-world network traffic datasets. We find that our solution outperforms the state-of-the-art (SOA) solutions of the conventional traffic classification \cite{lopez2017network} and CoS traffic classification \cite{d2021network}. To the best of our knowledge, this is the first work to employ time-distributed feature learning using deep learning for the NTC problem. The proposed solution can be a universal method to extract holistic-temporal features in any time-series classification task.



Our work spearheads the NTC problem for IoT networks to

\begin{itemize}
    \item develop a robust and initial raw-features independent time-distributed deep learning architecture for network traffic classification,
    \item introduces time-distributed feature learning to extract the holistic-temporal features from network traffic,
    \item and proposes a novel representation of network traffic flows as a greyscale video stream to enable time-distributed feature learning.
\end{itemize}
\color{black}

The rest of the paper is structured as follows. Section \ref{sec:rw} presents a comprehensive study of related works of network traffic classification. Section \ref{sec:prob_form} sets up the problem formulation. Section \ref{sec:TA} describes the theoretical analysis of the proposed solution to show the holistic-temporal feature extraction for network traffic classification. Section \ref{sec:EA} presents the implementation details of the models and the pre-processing of the data used for experiments. In Section \ref{sec:results}, we explain the experimental setup, compare the performance of the proposed solution with those of the SOA methods, and provide the time-distributed deep learning cost analysis. Finally, the paper is concluded in Section \ref{sec:conclusion} by providing the future directions of the research.

\section{Related Work}
\label{sec:rw}

Conventional NTC, also called multi-class classification, is essential for resource management, service level agreements and security monitoring in IoT networks. However, conventional NTC is expensive and laborious; therefore, the consequent delays might cause problems in scenarios where quick management is required for real-time tuning of IoT networks. To this end, CoS NTC is employed for such scenarios. CoS NTC identifies the type of service or traffic quickly; that information is usually sufficient to tune the network immediately. CoS has recently gained popularity because of the advent of demanding services for IoT devices. 

Deep learning models are promising solutions for NTC \cite{lopez2017network}. It is a supervised learning method that can find the best features for classification. Therefore, it avoids the manual feature selection step. However, deep learning structure is crucial in designing a successful classifier. The convolutional neural networks (CNNs), long-short-term-memory (LSTM), and sequential hybrid structure of CNN and LSTM in the form of CNN-LSTM are widely used in NTC \cite{lopez2017network,d2021network,manjunath2021time}. The main problem with the current deep learning classifier is that it has a tight dependency on initial hyperparameters and raw features. Therefore, \cite{d2021network} explores the autoencoder to remove the dependencies, and the work improves the classifiers’ performance. However, the autoencoder has not shown remarkable performance improvement for network traffic, which is time series data. In this section, we study conventional and CoS classifiers.

\subsection{Conventional Network Traffic Classifiers}

Lopez -Martin \textit{et al.} \cite{ lopez2017network } is the first work that employs deep learning for NTC. The authors have used CNN, LSTM, and CNN-LSTM for conventional traffic classification. The work uses the traffic extracted from redIRIS, a Spanish academic and research network. Six raw features, source port, destination port, packet direction, payload length, inter-arrival time, and window size, are used for $266,160$ network flows. CNN-LSTM is most successful in the experiment by achieving $96.32\%$ accuracy on the dataset used in the work.

A CNN-based classifier is used by Tong \textit{et al.} \cite{tong2018novel} and achieves 99.34\% of F1-score while classifying five applications using 20,000 flows data and 1,400 features. G. Aceto \textit{et al.} \cite{ aceto2018mobile, aceto2019mobile} use different deep learning models to classify mobile encrypted traffic. Each byte of the captured packet is treated as a feature. For uneven packets, padding with zero is added to normalize the data. The traffic from Android, iOS, and Facebook are captured for the project. The authors compare the performance of stacked autoencoders (SAE), two-dimensional CNN, one-dimensional CNN, LSTM, and two-dimensional CNN with LSTM in the classification problem. For the considered data set, 1D-CNN performed well with a $96.50\%$ accuracy.

Most traffic classifiers focus on application-level features; however, the work presented in \cite{9626148} tries to use radio-level features to avoid domain-based assumptions. The authors have generated data for the experiment. Two deep learning-based models are studied in the problem: CNN and gated recurrent neural networks (GRU). Six classes are classified in the multi-class problem. The solution achieves an accuracy of 91\% for conventional classifications.

A multi-scale feature attention-based CNN solution is proposed in \cite{9707490}. They analyze a specific pattern in network packet per flow for classification. Pure IoT device network traffic is classified in the work. This work uses initial byte-based raw features to develop feature maps to fuse with n-gram features to enable attention mechanisms. The solution achieves 94\% accuracy on one dataset and 98.8\% accuracy on another dataset.

A non-linear ensemble solution is proposed for traffic classification \cite{9841019}. Well-known machine learning algorithms are stacked at different stages to achieve a non-linear ensembling scheme to boost multi-class network traffic classification. Support vector machine (SVM), decision tree (DT), random forest (RF), multi-layer perceptron (MLP), AdaBoost, K-nearest neighbours (KNN), LightGBM, Catboost, and XGBoost are the few algorithms studied in the work. However, the authors propose an ensemble of DT, RF, AdaBoost, and XGBoost at two stages for classification. The proposed method uses two different datasets, one of which is also used in our experiments. Their ensemble method achieves 91\% on the dataset we used in our work.

Flow-level and packet-level characteristics are used in the data to classify encrypted network traffic \cite{10082962}. The work uses the hybrid neural network that uses packet length sequence as the flow-level feature and the initial byte of the packet as the packet-level feature. Two path analyses are done for packet length and packet bytes in hybrid neural networks using GRU and SAE, respectively. The output from GRU and SAE are fused to form final features at the decision layer. Their solution is tested on multiple datasets and achieves an average of 94.5\% accuracy.

\subsection{Class-of-Service Network Traffic Classifiers}

An unsupervised and supervised learning-based solution for CoS classification is proposed in \cite{Roughan2004class}. Interactive, bulk data transfer, streaming, and transactional types of services are classified in the work. Specifically, Skype call and Hangout call services are included in the interactive class. File transfer protocol and remote copy form the bulk data transfer. Different video services are used in the streaming class. Finally, chat services are called the transactional class. Simple linear discriminant analysis and KNN are employed, and they achieve 43\% and 86\% accuracy, respectively. 

The game theory-based solution is introduced in \cite{Chowdhury2019explaining} for CoS classification. Using the game theory, Shapley value-based super features are framed to represent the CoS traffic data. In the preprocessing, 266 raw features are used to find the Shapley value. The feed-forward neural network is employed in the work and achieves 92.5\%.

A dynamic classification procedure is proposed in \cite{aureli2020going} to detect distinctive traffic characteristics and assign them to different CoS traffic. The authors employ semi-unsupervised machine learning methodologies. Linear discriminant analysis and K-Means algorithms are used. With this solution, the authors identify Internetwork Control, Critical VoIP, Flash Voice, and other types of traffic. The solution achieves 91\% accuracy for the dataset used in the work.

The recent work in \cite{d2021network} proposes an auto-encoder-based deep learning architecture for the CoS classification type. The CNN-LSTM architecture achieves 97\% accuracy for CoS classification for four classes. 

Segmented learning using random forest is employed for CoS classification in \cite{manjunath2022segmented}. The solution tries to find statistical commonality among traffic segments for a successful classification. A heuristic-based technique is used to find the segment size and number of segments required for training. The solution tested uses two datasets and achieves 99.13\% for identifying seven different CoS such as File transfer, Video, VoIP, Remote cloud, virtual reality, and other services.

Online dynamic decentralized learning models such as RDOC-O and RDOC-C \cite{10287496} are proposed to classify malicious and non-malicious types of classes. Though the work is not directly related to classifying CoS traffic services, if we look at malicious and non-malicious traffic with subcategories, then the problem is similar to CoS traffic classification. Both malicious and benign traffic of VoIP, Chat, Web, FTP, Mail, and other services are classified. Their solution achieves an 85\% F1 score.

Based on our literature survey, the two main research gaps we identify are i) the need for a standard feature extraction method for both types of traffic classifications and ii) extracting only spatio-temporal features and ignoring the pseudo-temporal features. In addition to the above limitations, the existing works are tested on limited services and traffic flows. Therefore, the robustness and generalization of the solutions need further verification. The recent works \cite{ 9707490, 9626148 } use mostly CNN in their classification solutions. Only the works \cite{lopez2017network, d2021network} uses CNN-LSTM configuration. \cite{d2021network} is the only work to propose automatic feature selection using autoencoder. However, it is successful in extracting only spatiotemporal features. Since we consider CNN-LSTM and generic feature extraction in our work, we consider \cite{lopez2017network} and \cite{d2021network} as state-of-the-art solutions. Other than \cite{lopez2017network,d2021network}, the rest of the works in the literature review have used CNN, GRU, and other machine learning algorithms. Our work focuses on building a robust and generic holistic-temporal feature extraction for conventional and CoS classification.

\color{black}

Main notations are summarized in Table \ref{table_notation}.

\begin{table}[ht]
	\caption{Notation List}
	\label{table_notation}
	\centering
	\begin{tabular}{l p{5.5cm}}
		\toprule
		lowercase $z$&  scalar\\
		bold lowercase $\boldsymbol{z}$ & vector\\
		bold uppercase $\bm{Z}$ & matrix\\
        uppercase $Z$ & constant scalar\\
		$i$, $j$, $l$, $m$, $n$, $u$ & index of an element in a vector or column or row in a matrix \\
  $[\bm{Z}]_{i,:}$, $[\bm{Z}]_{:,j}$ & the $i$-th row and the $j$-th column of a matrix, respectively\\
		$[\bm{Z}]_{i:j,m:n}$ &sub-matrix with the $i$-th to the $j$-th rows and the $m$-th to the $n$-th columns\\
		$[\bm{Z}]_{i,j}$ &entry at the $i$-th row and the $j$-th column of a matrix\\
		$[\boldsymbol{z}]_{i}$ &the $i$-th element of a vector\\
  $[\boldsymbol{z}]_{i:j}$ &subvector with the $i$-th to the $j$-th elements\\
		$\sigma(\cdot)$ &  activation function\\
		\bottomrule
	\end{tabular}
\end{table}

\section{Problem Formulation}
\label{sec:prob_form}

The main aim of the work is to enable holistic temporal feature extraction for traffic classification in IoT networks. Therefore, we propose time-distributed feature learning using deep learning. 
\color{black}

In this section, we formulate the problem. Let the IoT traffic flows be given as:

\begin{equation}\label{datamat}
\bm{D} = [\bm{d}_1, \bm{d}_2,...,\bm{d}_m,..., \bm{d}_M]^T,
\end{equation}
where $\bm{d}_m \in \mathbb{R}^{N \times 1}$ ($m = {1,2,...,M}$) is the $m$-th flow with $N$ number of initial raw features. Every $m$-th flow in $\bm{D}$ has a true label $\mathcal{L} \in \{1,2,...,\mathcal{C}\}$, where $\mathcal{C}$ denotes the different traffic services.


The objective of employing deep learning is to map $\bm{d}_m$ traffic flow to one of the classes in $\mathcal{L}$. Let $\bm{p}_{\bm{d}_m}$ be the predicted probabilities for a flow $\bm{d}_m$ belonging to all the classes in $\mathcal{L}$. The final decision of a traffic flow belonging to one of the classes in ${\mathcal{L}}$ that consists of the most significant probability is given as:

\begin{equation}
\label{Equ:dl_E_1}
  \hat{{c}} = \text{arg} \max_{c} \left[\bm{p}_{\bm{d}_m}\right]_{c},
\end{equation}
where $[\cdot]_{c}$ is the $c$-th element of a vector, and $c \in \mathcal{L}$ is the index of the class.

Generally, deep learning uses initial features to find $\bm{p}_{\bm{d}_m}$. However, our solution needs to extract holistic-temporal features for a superior classification. We denote $\psi(\cdot)$ as a function that extracts the holistic-temporal features as given by:

\begin{equation}
\label{Equ:dl_E_2}
 \hat{\mathcal{Y}}_{\bm{d}_m} = \psi(\bm{d}_m),
\end{equation}
where, $\hat{\mathcal{Y}}_{\bm{d}_m}$ consists of holistic-temporal features for the flow $\bm{d}_m$ in $\bm{D}$.

We employ time-distributed deep learning to find $\psi(\cdot)$ for the network flows. We provide a mathematical explanation of the holistic-temporal feature extraction in the next section. In our solution, for the flow $\bm{d}_m$, the probability $\bm{p}_{\bm{d}_m}$ is found using the holistic-temporal features Eq. \eqref{Equ:dl_E_1} to make the final decision on which class it belongs to.

\section{Theoretical Analysis of
Time-distributed Feature Learning}
\label{sec:TA}

As we established, IoT traffic exhibits a highly random nature with stationarity and non-stationarity along the temporal aspect as the session progresses. Spatio-temporal features are insufficient; holistic temporal features can capture the IoT network traffic's randomness, enhancing performance. Time-distributed deep learning aids in extracting the holistic temporal features. IP and pseudo-spatial features derive spatial features extracted by CNN. Short and long-term temporal features are extracted by LSTM, and time-distributed deep learning extracts pseudo-temporal features. 

Holistic temporal features consist of intra-, inter-, and pseudo-temporal features. Intra-temporal features also consist of spatial features contributed by IP, which gives an idea of a local or wide-area network. Intra-temporal features are pseudo-spatial temporal features because of the data representation presented in section \ref{sec:DR}. Inter-temporal features are temporal information among the traffic flows. Pseudo-temporal features are the temporal dependencies extracted from inter- and intra-temporal information. Visualizing or explaining these features in deep learning is a complicated process since deep learning is a black box; however, we try to derive the general form of these features mathematically in this section. We also provide the required data representation, model descriptions, and mathematical explanations.

\subsection{Data Representation}
\label{sec:DR}

We use CNN in the initial stage of the time-distributed deep learning model. As we know, CNN expects input data to be images; therefore, we need to represent the network flow data in the image form. Recall Eq. \eqref{datamat} that there are $M$ samples of data with $N$ number of features, and then the whole data matrix $\bm{D}$ has a dimension of ${M\times N}$. Every $\bm{d}_{m}^T$ row of $\bm{D}$ should be represented as a new matrix with dimension $R\times C$, where $(R,C)$ is a factor pair of $N$, which is the length of $\bm{d}_{m}^T$. The problem formulation of the transformation is given by:

\begin{equation}
\label{Equ:DR_E_1}
\bm{d}_{m (N \times 1)} \rightarrow \bm{X}_{m (R \times C)},
\end{equation}
where $ m=1,2,\cdots, M$, and the subscript terms in the brackets represent the dimension of the vector or matrix. The newly represented matrix $\bm{X}_m$ is with a dimension $(R \times C)$. The mathematical transformation of \eqref{Equ:DR_E_1} is given by:

\begin{equation}
\label{Equ:DR_E_2}
 \left[\bm{X}_m\right]_{j,:} = [\bm{d}_m]_{\left(j-1\right)C + 1:jC},
\end{equation}
where $j = 1,2,...,R$, and the $j$-th row of a matrix is given by $[\cdot]_{j,:}$.

\begin{figure}
 	\centering
 	\includegraphics[trim={0 2 2 0},width=0.8\linewidth]{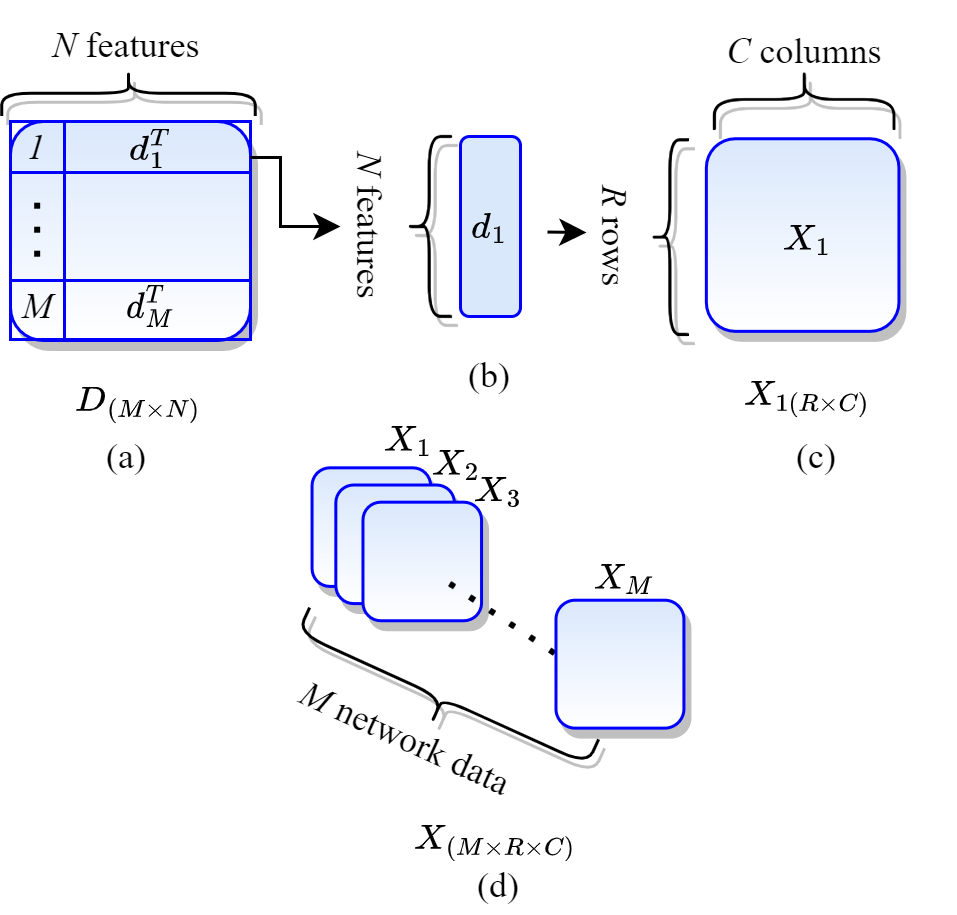} 
 	\caption{(a) The original raw data in matrix form with $M$ samples and $N$ features. (b) An $N \times 1$ column vector for one sample. (c) The transformed representation of the sample into a matrix format, and ($R,C$) is a factor pair of $N$. (d) The final representation of the grey scale video stream using Algorithm \ref{Al:DR_A_1}, and $\bar{\bm{X}}$ is an $M \times R \times C$ tensor.}
 	\label{fig:DR_F_1}
\end{figure}

Based on Eq. \eqref{Equ:DR_E_2}, we can transform a network flow of all the data to the two-dimensional (2D) representation, which can be seen as a video stream of grey-scale images. We summarize the representation technique in Algorithm \ref{Al:DR_A_1}. We can notice that we still retain the linear relation of the basic features and make it possible to find more patterns along the neighbour data. The algorithm's output is a tensor $\bar{\bm{X}}$ with dimension ($M \times R \times C$). The pictorial view of the representation is shown in Fig. \ref{fig:DR_F_1}.

\color{black}
\begin{algorithm}
\caption{Network flows representation as a grey-scale video stream}\label{Al:DR_A_1}

\KwData{Input: $\boldsymbol{D}$}
\KwResult{Output: $\bar{\bm{X}}$}
 initialization\;
 
    Choose $R$ and $C$ that are factors of  the length of $\boldsymbol{d}_m$
 
 \While{$m \leqslant M$}{ 
    \While{$j \leqslant R$}{
        $\left[\bm{X}_m\right]_{j,:}=\left[\boldsymbol{d}_m\right]_{(j-1)C+1:jC}$
    }
    $\bar{\bm{X}}$.append($\bm{X}_m$)
  }
\Return $\bar{\bm{X}}$
\end{algorithm}

For an LSTM, we use the data in the original representation $\bm{d}_{m}$, as given by Fig. \ref{fig:DR_F_1}(a) and Fig. \ref{fig:DR_F_1}(b).
\color{black}

\subsection{Model Description}
\label{sec:math_model}

\begin{figure}
 	\centering
 	\includegraphics[width=\linewidth]{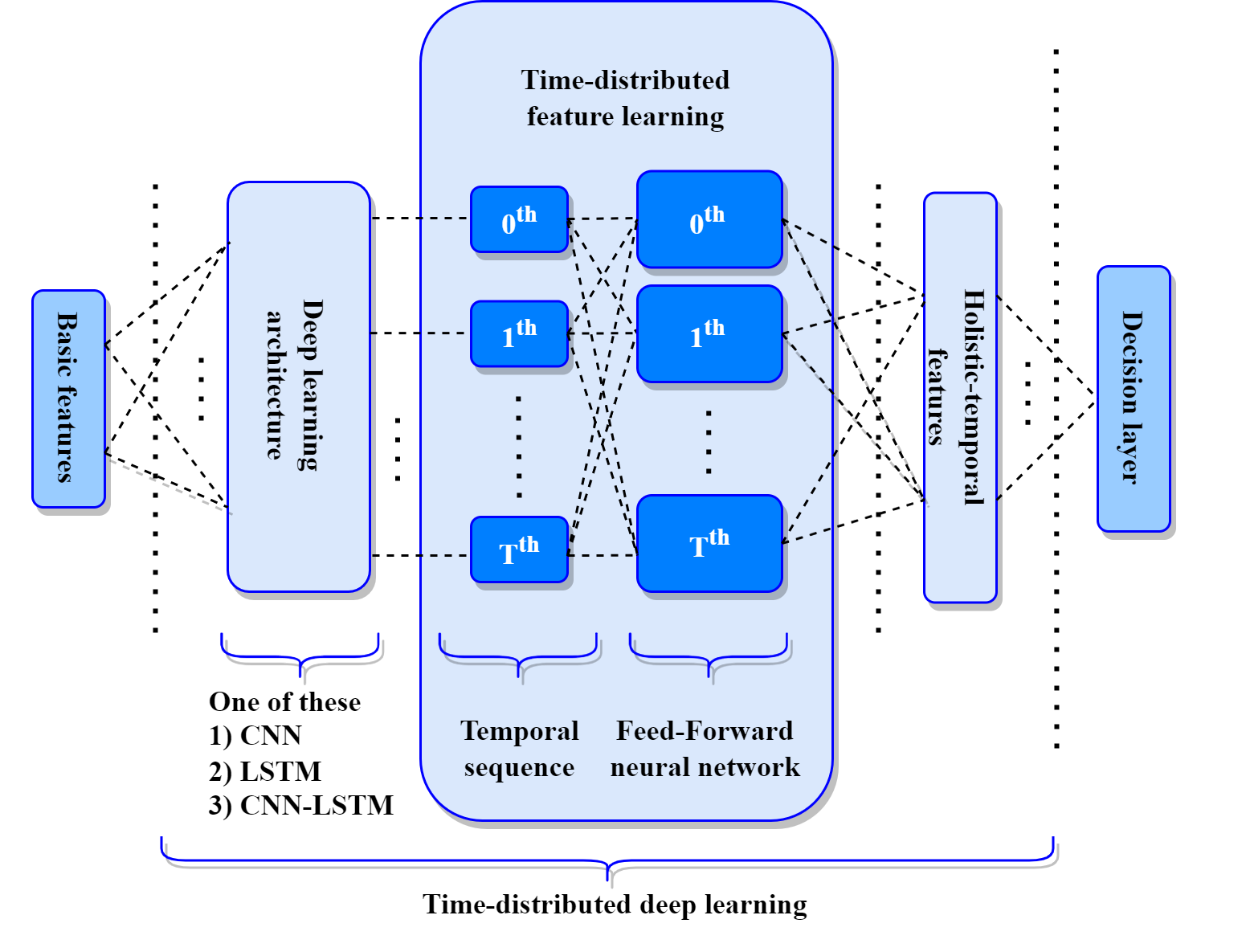} 
 	\caption{Holistic-temporal feature extraction using time-distributed feature learning employing deep-learning. We use CNN, LSTM, or CNN-LSTM as the deep learning architecture. Holistic-temporal features extracted by time-distributed deep learning are fed to the decision layer. The CNN-TD(FFNN) is expected to extract only inter-temporal and pseudo-temporal features. The LSTM-TD(FFNN) extracts inter and pseudo-temporal features. The CNN-LSTM-TD(FFNN) extracts the holistic, i.e., intra, inter, and pseudo-temporal features.}
 	\label{fig:MD_F_1}
\end{figure}

Fig. \ref{fig:MD_F_1} shows our proposed deep-learning-based time-distributed feature learning for IoT NTC. CNN is well-known for its ability to extract spatial or intra-temporal features in IoT network traffic \cite{khan2019energy}. Likewise, LSTM is known to identify the temporal patterns or inter-temporal features among the network flows \cite{rezaei2019deep}. Therefore, in the deep learning layer, we use one of the three structures from i) CNN, ii) LSTM, and iii) CNN-LSTM. We add the time-distributed layer after the deep learning layer, explicitly applying a feed-forward neural network (FFNN) wrapped in a time-distributed wrapper to extract pseudo-temporal features. The time-distributed structure enables a parallel feature extraction from every temporal sequence extracted by previous layers.  The holistic-temporal features extracted by time-distributed deep learning are used in classification in the final decision layer.

Based on the three different deep-learning architectures as shown by Fig. \ref{fig:MD_F_1}, we study the following three new models,
\begin{itemize}
    \item Model 1: CNN-TD(FFNN) consists of CNN and a time-distributed feed-forward neural network.
    \item Model 2: LSTM-TD(FFNN) consists of LSTM and a time-distributed feed-forward neural network.
    \item Model 3: CNN-LSTM-TD(FFNN) consists of CNN-LSTM and a time-distributed feed-forward neural network.
\end{itemize}

We also compare the vanilla versions of the above three models, in which the time-distributed layer is removed. In the following section, we derive the mathematical expression for all models to find the superiority of the time-distributed feature learning. The classification performance is validated empirically. 

\subsection{Model 1: CNN-TD(FFNN)}
\label{sec:CNN}
We present the mathematical representation of the features extracted by our models. For CNN, the input is $\bm{X}_m$, $m = 1,2,\cdots,M$, as given by Fig. \ref{fig:DR_F_1}(d). For simplicity, let's consider a single flow, denoted as $\bm{X}$, without any subscript.

The CNN layer performs the convolution operation between the input $\bm{X}$ and kernel windows. The result of the convolution operation, $\bm{Y}$, is given as:

\begin{gather}
\label{Equ:mod1_E_1}
 [\bm{Y}]_{i,j} = \sum_{l=1}^{L}\sum_{p=1}^{P}\sum_{q=1}^{Q}\left[\bm{K}_l\right]_{p,q} \left[\bm{X}\right]_{i+p-1,j+q-1},\\
 i=1,\cdots,Y_\text{row}, \; j=1,\cdots,Y_\text{col}, \nonumber
\end{gather}
where $\bm{K}_l \in \mathbb{R}^{P \times Q}$ is the $l$-th kernel window, $P$ and $Q$ are the sizes of the kernel window, ${Y}_\text{row}$ is the number of rows and ${Y}_\text{col}$ is the number of columns of $\bm{Y}_m$, respectively, as given by:
\begin{equation}
\label{Equ:mod1_E_2}
 {Y}_{row} = \frac{R - P + 2G}{S_x} + 1,
\end{equation}
\begin{equation}
\label{Equ:mod1_E_3}
{Y}_{col} = \frac{C - Q + 2G}{S_y} + 1,
\end{equation}
in which $S_x$ and $S_y$ are the row and column strides of the moving kernel, respectively, and $G$ is the number of padding.

The intra-temporal features extracted by one CNN unit are given by Eq. \eqref{Equ:mod1_E_1}. We feed Eq. \eqref{Equ:mod1_E_1} to a $2 \times 2$ maxpool and batch normalization (BN) layers to obtain $\tilde{\bm{y}}_\text{C}$. The output after BN exists in a tensor form, which needs to be flattened to have $\tilde{\bm{y}}_\text{C}$. We can have $U$ outputs, from $U$ CNN units, denoted as $\tilde{\bm{y}}_{\text{C}u}$, $u=1,2,\cdots,U$. Every extracted feature is then fed to time-distributed FFNN to have $\hat{\mathcal{Y}}_\text{C}$, which is given as:
\begin{equation}
\label{Equ:mod1_E_4}
 \hat{\mathcal{Y}}_\text{C} = \sum_{u=1}^{U} \sigma\left(\bm{W}_{\text{C}u}  \tilde{\bm{y}}_{\text{C}u}+ \bm{b}_{\text{C}u}\right),
 \end{equation}
where $\bm{W}_{\text{C}u}$ is the weight matrix, $\bm{b}_{\text{C}u}$ is the bias vector, $\sigma(\cdot)$ is the ReLU activation function.



The CNN's $\bm{Y}$ output consists of intra-temporal features. Upon applying a time-distributed wrapper, we obtain Eq. \eqref{Equ:mod1_E_4}, which consists of intra (given by $\tilde{\bm{y}}_{\text{C}u}$ in Eq. \eqref{Equ:mod1_E_4})  and pseudo-temporal (given by outer summation in Eq. \eqref{Equ:mod1_E_4}) features. We can use a decision-making FFNN with a Softmax to find traffic services by feeding Eq. \eqref{Equ:mod1_E_4}.

For comparative analysis, let's consider the output of CNN without a time-distributed layer, which is given by:
\begin{equation}
\label{Equ:mod1_E_6}
 \hat{\mathcal{Y}}_{\text{CNN}} = \sigma(\bm{W}_\text{C}\tilde{\bm{y}}_\text{C} + \bm{b}_\text{C}),
\end{equation}
where $\bm{W}_\text{C}$, $\tilde{\bm{y}}_\text{C}$, and $\bm{b}_\text{C}$ are the weight, flattened output and bias, respectively. The pseudo-temporal features from Eq. \eqref{Equ:mod1_E_4}, presented by outer summation, are absent in Eq.\eqref{Equ:mod1_E_6}. Thus, Eq. \eqref{Equ:mod1_E_6} gives only intra-temporal features.


\subsection{Model 2: LSTM-TD(FFNN)}
\label{sec:LSTM}

The data in the original format shown in Eq. \eqref{datamat} are used directly for the LSTM-based model having LSTM in the first layer. Gates in the LSTM cell update recursively for all input and give an intermediate output vector $\tilde{\bm{y}}_\text{L}$ with a dimension depending on the number of the LSTM cells used in the model. Refer to \cite{d2021network} for the LSTM gate models. The output of an LSTM layer is given as:

\begin{equation}
\label{Equ:mod2_E_7}
\tilde{\bm{y}}_{\text{L}} = \bm{D}^T \bm{\phi},
\end{equation}
where $\bm{\phi}$ is a non-linear multi-variable vector-based function with the short and long dependencies from an arbitrary LSTM cell. The approximation of $\bm{\phi}$ can be found in \cite{d2021network}.

The output dimension of an LSTM layer for a data sample $\bm{D}$ with $N$ number of features is $N \times 1 \times 1$. To employ time-distributed feature learning, we extract the input features for every data sample to form a pseudo-temporal sequence. Therefore, we apply $N$ time-distributed FFNN units and the output before the final decision-making layer is given as:
\begin{equation}
\label{Equ:mod2_E_8}
 \hat{\mathcal{Y}}_{\text{L}} = \sum_{n=1}^{N} \sigma\left(\bm{W}_{\text{L}n}\tilde{\bm{y}}_{\text{L}} + \bm{b}_{\text{L}n}\right),
 \end{equation}
where $\bm{W}_{\text{L}n}$ is the weight matrix, and $\bm{b}_{\text{L}n}$ is the bias vector.

In Eq. \eqref{Equ:mod2_E_8}, $\tilde{\bm{y}}_\text{L}$ captures all possible short and long inter-temporal features, and the outer summation extracts all possible pseudo-temporal features. 

Similar to what we do at the end of Section \ref{sec:CNN}, we compare the output of the time-distributed variant given by Eq. \eqref{Equ:mod2_E_8} with that of the LSTM-FFNN without the time-distributed layer, which is given by:

\begin{equation}
\label{Equ:mod2_E_9}
 \hat{\mathcal{Y}}_\text{{LSTM}} =  \sigma(\bm{W}_\text{L} \tilde{\bm{y}}_\text{L} + \bm{b}_\text{L}),
 \end{equation}
where $\bm{W}_\text{L}$, $\tilde{\bm{y}}_\text{L}$, and $\bm{b}_\text{L}$ are the weight, LSTM output and bias, respectively.
We can see that Eq. \eqref{Equ:mod2_E_9} lacks the outer summation introduced by the time-distributed layer and thus is not able to extract the pseudo-temporal features.

\subsection{Model 3: CNN-LSTM-TD(FFNN)}
\label{sec:model3}

We have shown the intra-temporal and inter-temporal feature extraction using CNN and LSTM in Models 1 and 2, respectively. In Model 3, we use CNN-LSTM to derive the advantages of extracting intra- and inter-temporal features, followed by a time-distributed FFNN to extract pseudo-temporal features. As a result, Model 3 extracts the holistic temporal features proposed in the paper.


The CNN's output is given in Eq. \eqref{Equ:mod1_E_1} for a single unit. However, the final output of $U$ CNN units is given by $\tilde{\bm{y}}_{\text{C}u}$. The tensor form of CNN's output is reshaped to a matrix form given as $\tilde{\bm{Y}}_{\text{C}u}$, which is necessary for the pipeline of CNN-LSTM structure. Practically, we can achieve the form using the reshape API from sklearn \cite{scikit-learn}. The final output after the LSTM layer is given as:
\begin{equation}
\label{Equ:mod3_E_7}
\tilde{\bm{y}}_{\text{CL}} = \tilde{\bm{Y}}_{\text{C}u}^T \bm{\phi}.
\end{equation}

Finally, we employ the time-distributed layer to every temporal sequence from the output of the LSTM layer. 
Assume the CNN layer extracts $T$ number of intra-temporal features, and if we use $K$ LSTM cells, the dimension of $\tilde{\bm{Y}}_{\text{C}u}$ is $T \times K \times 1$. Then, we need $K$ time-distributed units to extract the pseudo-temporal features, as given by:
\begin{equation}
\label{Equ:mod3_E_8}
 \hat{\mathcal{Y}}_{\text{CL}} = \sum_{t=1}^{T} \sigma\left(\bm{W}_{\text{CL}t}\tilde{\bm{y}}_{\text{CL}} + \bm{b}_{\text{CL}t}\right),
\end{equation}
where $\bm{W}_{\text{CL}t}$ is the weight matrix and $\bm{b}_{\text{CL}t}$ is the bias.

The final feature extraction consists of the holistic-temporal features. Specifically, in Eq. \eqref{Equ:mod3_E_8}, $\tilde{\bm{y}}_{\text{CL}}$ consists of intra and inter-temporal features extracted by CNN-LSTM. The outer summation extracts the pseudo-temporal relations to contribute to holistic features that can provide greater information for an IoT traffic classification. In the following section, we show the efficiency of these features in experimental classification.

\section{Model Implementation and Data Pre-processing}
\label{sec:EA}


In this section, we explain the implementation of proposed models and pre-processing of the data.

\subsection{Implementation Details of Models}

\begin{figure}
 	\centering
 	\includegraphics[width=0.9\linewidth]{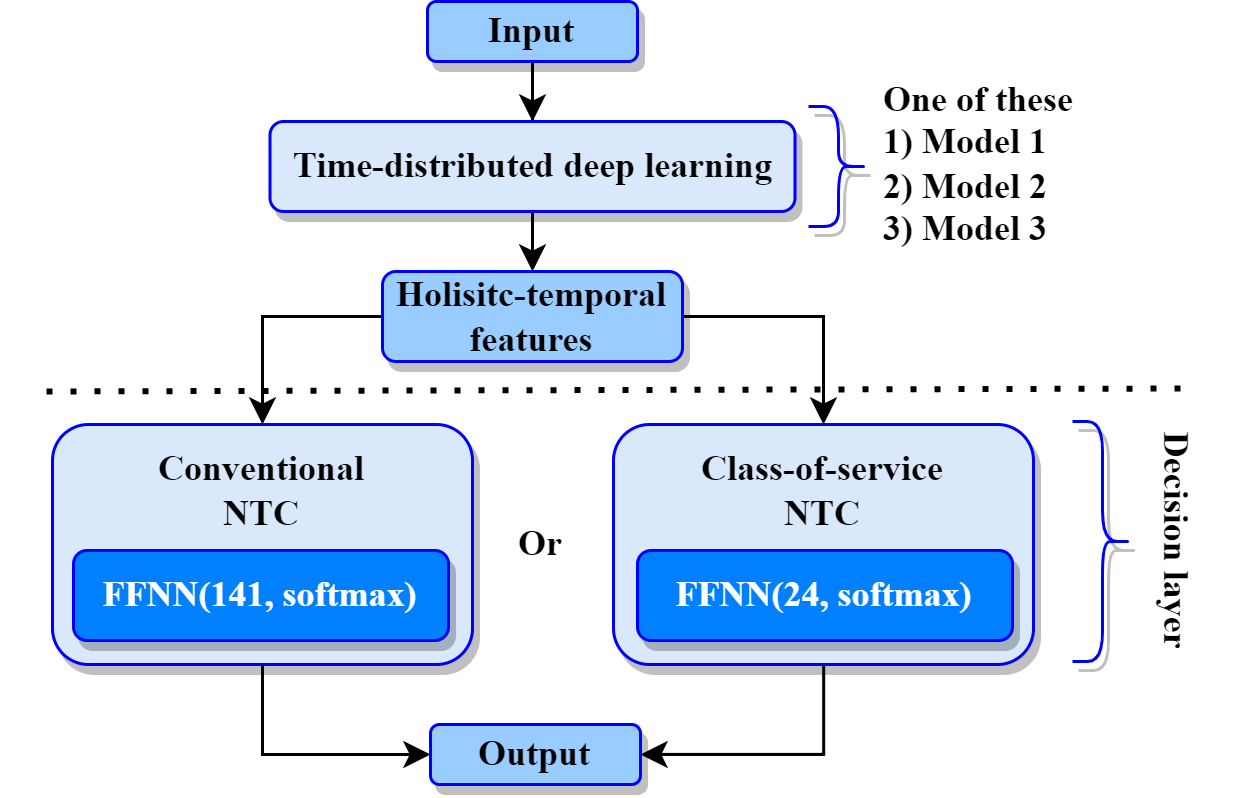} 
 	\caption{Holistic-temporal feature extraction from time-distributed deep learning as input for Decision layer for conventional or CoS classification. The Decision layer is implemented with an FFNN and softmax activation, and therefore, the output is a probability distribution, and the final decision is made by Eq. \eqref{Equ:dl_E_1}. Time-distributed deep learning can be any of the 3 model configurations. The time-distributed deep learning layer consists of a deep learning layer and a time-distributed layer from Fig. \ref{fig:MD_F_1}. The exact number of units for the decision layer FFNN for all datasets is given in Table \ref{table:dataset}.
  }
 	\label{fig:DOM_F_1}
\end{figure}

The general implementation flow of the holistic-temporal feature extraction is given in Fig. \ref{fig:DOM_F_1}. The deep-learning layer takes a different form based on the models. As we established, we study three different deep-learning structures. We expect holistic-temporal features after the time-distributed deep-learning layer, which are used for classification tasks. For the Unicauca Network Flows Dataset \cite{kaggle}, in the final decision layer, we have to use 141 units for multi-class classification and 24 units for CoS classification. The number of FFNN units in the Decision layer depends on the number of classes $\mathcal{C}$ defined in the problem, as given by Section \ref{sec:prob_form}. More details on data can be found in Section \ref{Datapreproc}. The output is collected in the form of a probability distribution. The class with the highest probability is the final decision for a given sample using Eq.\eqref{Equ:dl_E_1}.

\subsubsection{Model 1}
\begin{figure}
 	\centering
 	\includegraphics[width=0.8\linewidth]{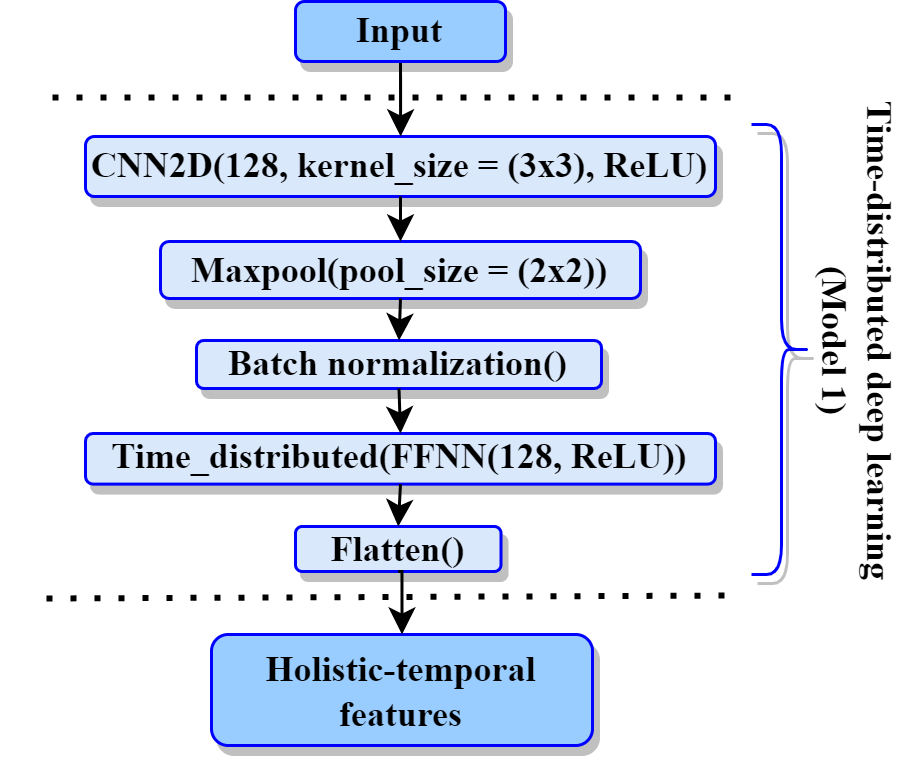} 
 	\caption{Implementation details of Model 1. 128 units of 2D CNN are used with a 3$\times$3 kernel window. $2 \times 2$ maxpool is introduced between CNN2D and batch normalization layers. The output of batch normalization consists of intra- or spatio-temporal features and is fed to 128 units of time-distributed FFNN. The extracted features consist of intra- and pseudo-temporal features.}
 	\label{fig:DOM_F_2}
\end{figure}

Model 1 uses a 2D convolution layer with 128 units followed by 2D max-pooling and batch normalization. The time-distributed layer is employed for every temporal sequence extracted by the prior layers. The output of the time-distributed layer is flattened to feed it as an input to the Decision layer. The Model 1 implementation without the Decision layer is shown layer-wisely in Fig. \ref{fig:DOM_F_2}.

\subsubsection{Model 2}

\begin{figure}
 	\centering
 	\includegraphics[width=0.8\linewidth]{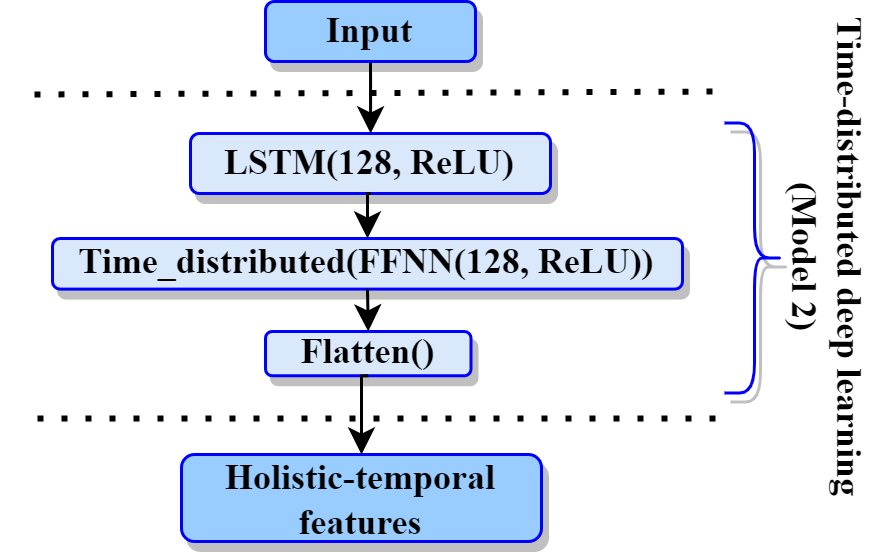} 
 	\caption{Implementation details of Model 2. 128 units of LSTM are employed in the first layer with ReLU activation. The initial layer of LSTM extracts the inter-temporal features, which are fed to the time-distributed FFNN of 128 units. The flattened final output consists of inter- and pseudo-temporal features.}
 	\label{fig:DOM_F_3}
\end{figure}

The LSTM layer with 128 units is used in Model 2. The time-distributed layer is applied for all time-sequenced data extracted by the LSTM layer. The output of the time-distributed layer is flattened, which is nothing but the holistic-temporal features. The Model 2 implementation is shown layer-wisely in Fig. \ref{fig:DOM_F_3}.

\subsubsection{Model 3}

\begin{figure}
 	\centering
 	\includegraphics[width=0.8\linewidth]{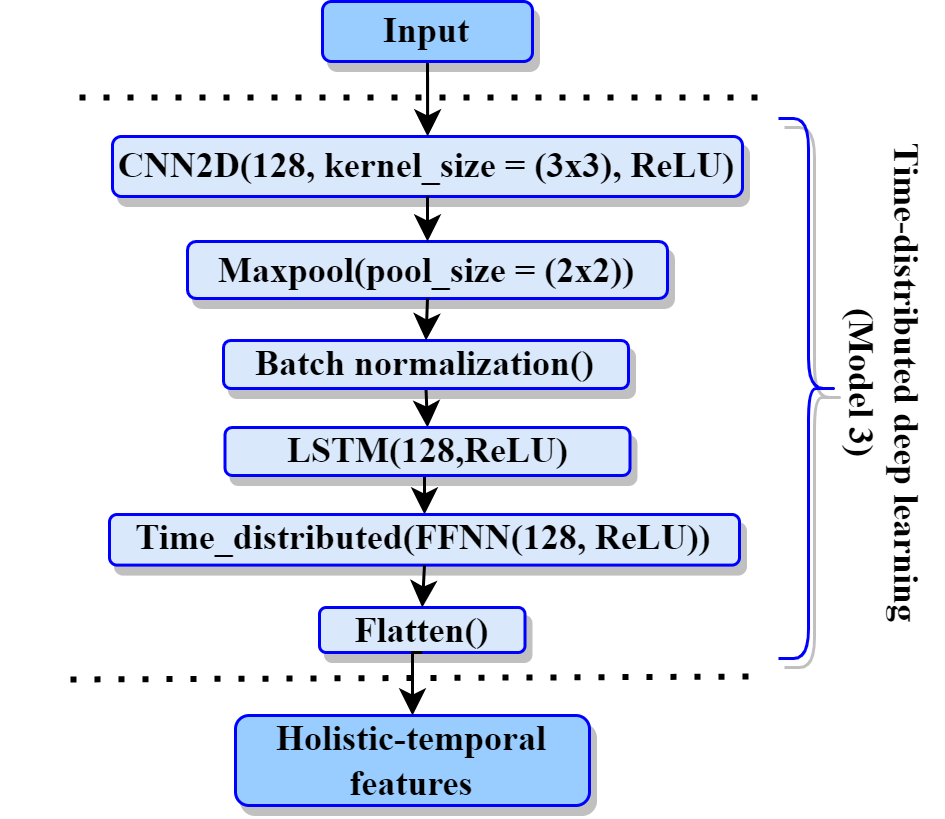} 
 	\caption{Implementation details of Model 3. Model 3 uses CNN2D and LSTM along with time-distribution FFNN. We use CNN2D as the first layer to utilize the advantage of high-dimensional data and to reduce the data dimension for future layers. It extracts the spatio-temporal features at the beginning to improve performance. The spatio-temporal data extracted by CNN is fed to LSTM of 128 units. LSTM helps to extract the interflow features. Finally, the time-distributed FFNN extracts the pseudo-temporal features. The flattened data after time-distributed FFNN is nothing but holistic-temporal features.}
 	\label{fig:DOM_F_4}
\end{figure}

Model 3 employs a 2D CNN with 128 units followed by 2D maxpool and batch normalization layers. An LSTM layer with 128 units is added sequentially. Then, we apply a time-distributed layer for every sequence of temporal feature extraction from the CNN-LSTM layer. The flattened output of the time-distributed layer is input to the Decision layer of FFNN with the softmax activation function. The Model 3 implementation is shown in Fig. \ref{fig:DOM_F_4}.

\subsection{Dataset and Pre-processing}
\label{Datapreproc}


We use multiple datasets to validate our solution. As we established, IoT traffic is a kind of IP traffic; however, IoT traffic exhibits temporally ranodm behaviour at higher rates of data transaction. Though the initial intention of the solution is to validate the IoT traffic using time-distributed feature learning, we wish to generalize the solution for different types of network traffic datasets. Therefore, in Dataset I, we include IP network traffic that depicts the IoT traffic characteristics \cite{kaggle}. Dataset II includes encrypted IP traffic \cite{aer}. Dataset III consists of flows of attacks in the IoT network \cite{s23135941}. To study the scalability and future compatibility of the solution, we form Dataset IV by combining an old dataset consisting of different application protocols, VPN and non-VPN datasets \cite{ vpn_non_vpn }, and the latest Virtual Reality network traffic dataset \cite{9685808}, a recent and hot-trend application.

Time-distributed feature learning is independent of the initial features; however, it depends on the data representation in section \ref{sec:DR}. The deep learning model extracts time-distributed features, including intra, inter, and pseudo-temporal features, which is a black box; however, a mathematical explanation behind the time-distributed features is given in section \ref{sec:math_model}. In this section, we discuss the raw features used in the work.

Except for the IoT traffic \cite{s23135941}, all other datasets are in packet capture (pcap) format. The pcap files are fed to the flow labeller \cite{FlowLabeler}, which uses the nDPI library \cite{deri2014ndpi} to extract the flow statistical features of the network traffic. The filename provides the label that is added to the extracted feature set. This process converts the dataset into comma-separated value (CSV) files. Artificial intelligence libraries implemented in Python can use CSV files in training, validation, and testing processes. The extracted CSV consists of 48 features, including source/destination IP and source/destination port, statistical features such as minimum, average, and maximum inter-arrival time, and the label. Please refer to \cite{kaggle} and \cite{FlowLabeler} for the features' details. The given 48 features are the same across the dataset except for the IoT traffic \cite{s23135941}; the dataset providers use their custom method to extract the statistical features of network attack traffic in the IoT setup. The details of the features for \cite{s23135941} can be found in \cite{CiCIoTDataset}.  The details of the datasets used in the work are given in Table \ref{table:dataset}.

\begin{table*}
\caption{Details of the datasets used to validate the time-distributed feature learning solution}
\centering
\begin{tabular}{
>{\centering\arraybackslash}m{1.1cm}  
>{\centering\arraybackslash}m{4.5cm}  
>{\centering\arraybackslash}m{1.51cm} 
>{\centering\arraybackslash}m{1.5cm}  
>{\centering\arraybackslash}m{1.5cm}  
>{\centering\arraybackslash}m{2cm}    
>{\centering\arraybackslash}m{2cm}    
}
\hline
&{Dataset} & {Number of flows} & {Number of features} & {IoT setup} & {Number classes for multi-class NTC} & {Number classes for CoS NTC} \\
\hline
Dataset 1 & Unicauca Network Flows Dataset \cite{kaggle}  & 2,704,839 & 48 & Yes & 141 & 24 \\
\hline
Dataset 2 & Encrypted mobile instant messaging traffic dataset \cite{aer}  & 54,458 & 48 & Yes & 10 & 8 \\
\hline
Dataset 3 & CIC IoT dataset \cite{CiCIoTDataset}  & 2,366,956 & 46 & Yes & 34 & 2/7 \\
\hline
Dataset 4 & VPN-nonVPN dataset \cite{VPNNONVPN} + VR gaming dataset \cite{9685808} & 6,237 & 48 & No & 15 & 2 \\
\hline
\end{tabular}

\begin{tablenotes}
\item Note: We define two types of CoS problems using the CIC IoT dataset, Dataset III. The dataset provides two classes (malicious and benign) and seven classes of different types of attacks. Detailed information about the classes can be found at \cite{CiCIoTDataset}. Dataset IV is formed by combining the VPN-nonVPN dataset \cite{VPNNONVPN}, which provides 14 different classes, and we add VR service data from \cite{9685808}. Therefore, in total, we have 15 classes. Dataset IV is formed to study the future compatibility of the solution for new applications. Datasets captured using multiple IoT devices in the capture environment are assumed to have an IoT setup. Datasets 1, 2, and 3 show IoT setup.
\end{tablenotes}

\label{table:dataset}
\end{table*}

We normalize the raw data to avoid bias. Features in string format are encoded to numbers using LabelEncoder API from sklearn \cite{scikit-learn}. Then, we employ the min-max scaling \cite{patro2015normalization} with a scale of [0,1], i.e., the minimum and maximum values of a feature are assigned with a value of 0 and 1, respectively, and every other value is scaled within the range of [0,1]. In practice, we use the sci-kit learn library \cite{scikit-learn}, which provides a convenient software application interface, to implement the min-max scaling and normalization of the data. For Model 1 and Model 3, we represent the scaled data in the image form as explained in Section \ref{sec:DR}, and we use the scaled data directly for Model 2. We use 70\% of the data for training, 10\% of the data for validation, and 20\% of the data to test the model.

\color{black}

\section{Experimental Evaluation}
\label{sec:results}

We study the solution empirically through different experiments in the following sub-sections. To study the effectiveness of the time-distribution solution, we consider all three models with and without the time-distribution layer presented in the previous sections. We also implement the SOA solutions of conventional NTC \cite{lopez2017network} and CoS NTC \cite{d2021network}. We use the data introduced in Section \ref{Datapreproc} to train and test the SOA models for a fair comparison since the datasets used in those works are unavailable. All experiments are conducted on a system with a graphical processing unit of Nvidia RTX 2080 Super with 8 GB memory. Required software modules are implemented in Python using TensorFlow \cite{abadi2016tensorflow}, Keras \cite{gulli2017deep}, and sklearn \cite{scikit-learn}.

\subsection{Performance Metrics}
We compute the performance metrics using the predicted classes and the ground truth labels. Specifically, we quantify the performance of the classification models using \textit{Accuracy}, \textit{Precision}, \textit{Recall}, and \textit{F}1-score as given by
\begin{equation}
 \label{Equ:Accuracy}
 \textit{Accuracy} = \frac{\text{TP}+\text{TN}}{\text{TP}+\text{TN}+\text{FP}+\text{FN}},
 \end{equation}
\begin{equation}
 \label{Equ:Precision}
 \textit{Precision} = \frac{\text{TP}}{\text{TP}+\text{FP}},
\end{equation}
\begin{equation}
 \label{Equ:Recall}
 \textit{Recall} = \frac{\text{TP}}{\text{TP}+\text{FN}},
 \end{equation}
 \begin{equation}
 \label{Equ:F1}
 \textit{F}1 = 2\times \frac{\textit{Precision}\times \textit{Recall}}{\textit{Precision}+\textit{Recall}},
 \end{equation}
where false-positive (FP) represents the false detection of the flows when the flow is not present in reality, false-negative (FN) represents no detection when the flows exist in reality, true-negative (TN) is the correct no-detection, and true-positive (TP), on the contrary, represents the correct detection. We also study the total time to finish the training.

The \textit{Accuracy} metric provides how many test samples are classified correctly. \textit{Precision} explains the number of test samples correctly predicted in the cases where the samples were positive. \textit{Recall} indicates how many actual positive cases are predicted correctly using our model. \textit{F}1-score is the harmonic mean of \textit{Precision} and \textit{Recall}. We have a higher \textit{F}1-score when a model achieves balanced \textit{Precision} and \textit{Recall}. Therefore, a model with a higher \textit{F}1-score is always better in the classification. In addition, we evaluate the total training time from the Keras application interface.

\subsection{Comparison with Vanilla Models without Time-distributed Learning}
\label{sec:comparisonvanilla}

We first investigate the performance of the proposed three new models and compare it with that of their vanilla counterparts without time-distributed learning, i.e., Model 1 vs. CNN-FFNN, Model 2 vs. LSTM-FFNN, and Model 3 vs. CNN-LSTM-FFNN. We train each model five times using random hyper-parameters, and the model's average training time helps decipher the convergence robustness and consistency of the model performance. In this section, we provide the detailed results for all models using the Dataset 1, Unicauca Network Flows Dataset \cite{kaggle}. For other datasets, we provide the average results for model 3 for vanilla and time-distribution variants for multi-class and CoS NTC in section\ref{othdataset}.  

\subsubsection{Model 1}

\begin{table}
\begin{threeparttable}
\caption{Performance Metrics and Training Time for Model 1.}
\centering 
\begin{tabular}
{
>{\centering\arraybackslash}m{0.9cm} 
>{\centering\arraybackslash}m{0.5cm} 
>{\centering\arraybackslash}m{0.8cm} >{\centering\arraybackslash}m{0.8cm} >{\centering\arraybackslash}m{0.8cm}
>{\centering\arraybackslash}m{0.8cm}
>{\centering\arraybackslash}m{0.8cm}
>{\centering\arraybackslash}m{0.5cm}
}
\hline
\multirow{7}{0.9cm}{\centering{Conven-\\tional}} &
Trial & Accuracy & Precision & Recall & F1 & Time (min) \\
\cline{2-7}
&\textbf{1}      & \textbf{0.901} & \textbf{0.897}  & \textbf{0.901} & \textbf{0.895}                & \textbf{3.90}        \\
&2      & 0.901 & 0.897  & 0.901 & 0.894                & 3.45       \\
&3      & 0.881 & 0.879  & 0.881 & 0.870                 & 1.50        \\
&4      & 0.889 & 0.887 & 0.889 & 0.880                 & 2.10        \\
&5      & 0.896 & 0.893  & 0.896 & 0.889                 & 3.30        \\
\cline{2-7}
&\textbf{Avg.}   & \textbf{0.894} & \textbf{0.891} & \textbf{0.894} & \textbf{0.885}                & \textbf{2.85}      \\
\hline
\multirow{7}{0.9cm}{\centering{CoS}} &
Trial & Accuracy & Precision & Recall & F1 & Time (min) \\
\cline{2-7}
&1      & 0.966 & 0.965  & 0.966 & 0.964                 & 0.93       \\
&2      & 0.969 & 0.969  & 0.969 & 0.968                  & 1.06       \\
&3      & 0.969 & 0.969  & 0.969 & 0.967                  & 1.06       \\
&4      & 0.974 & 0.974  & 0.974 & 0.972                 & 1.73       \\
&\textbf{5}      & \textbf{0.974} & \textbf{0.974}  & \textbf{0.974} & \textbf{0.973}                 & \textbf{1.33}       \\
\cline{2-7}
&\textbf{Avg.}   & \textbf{0.971} & \textbf{0.970} & \textbf{0.971} & \textbf{0.969}              & \textbf{1.22}\\
\hline
\end{tabular}
\begin{tablenotes}[para,flushleft]
Note: The performance metrics given are on the test dataset. The training time is the time taken by Model 1 during training. There are fewer classes in the CoS problem compared to the conventional NTC problem, and therefore a shorter training time.
\end{tablenotes}
\label{table:Performance metrics and training time for Model 1 with time-distributed feature learning.}
\end{threeparttable}
\end{table}

\begin{table}
\begin{threeparttable}
\caption{Performance Metrics and Training Time for the Vanilla Version of Model 1 without Time-distributed Feature Learning.}
\centering 
\begin{tabular}
{
>{\centering\arraybackslash}m{0.9cm} 
>{\centering\arraybackslash}m{0.5cm} 
>{\centering\arraybackslash}m{0.8cm} >{\centering\arraybackslash}m{0.8cm} >{\centering\arraybackslash}m{0.8cm}
>{\centering\arraybackslash}m{0.8cm}
>{\centering\arraybackslash}m{0.8cm}
>{\centering\arraybackslash}m{0.5cm}
}
\hline
\multirow{7}{0.9cm}{\centering{Conven-\\tional}} &
Trial & Accuracy & Precision & Recall & F1 & Time (min) \\
\cline{2-7}
&1      & 0.878 & 0.875  & 0.878 & 0.871                & 1.35       \\
&2      & 0.839 & 0.835  & 0.839 & 0.832                 & 1.35       \\
&\textbf{3}      & \textbf{0.889} & \textbf{0.887}  & \textbf{0.889} & \textbf{0.880}                & \textbf{2.25}       \\
&4      & 0.775& 0.774  & 0.775 & 0.764                & 1.20        \\
&5      & 0.855 & 0.861  & 0.855 & 0.858                 & 1.50        \\
\cline{2-7}
&\textbf{Avg.}   & \textbf{0.847} & \textbf{0.846}  & \textbf{0.847} & \textbf{0.841}               & \textbf{1.53} \\
\hline
\multirow{7}{0.9cm}{\centering{CoS}} &
Trial & Accuracy & Precision & Recall & F1 & Time (min) \\
\cline{2-7}
&1      & 0.780 & 0.799  & 0.780 & 0.755                 & 0.46       \\
&2      & 0.814 & 0.840  & 0.814 & 0.833                 & 0.80        \\
&\textbf{3}      & \textbf{0.851} & \textbf{0.871}  & \textbf{0.851} & {0.843}                 & \textbf{0.70}      \\
&4      & 0.855 & 0.878  & 0.855 & 0.840                 & 0.93       \\
&5      & 0.844 & 0.828  & 0.844 & 0.818                 & 1.20        \\
\cline{2-7}
&\textbf{Avg.}   & \textbf{0.829} & \textbf{0.843}  & \textbf{0.829} & \textbf{0.818}               & \textbf{0.81} \\
\hline
\end{tabular}
\begin{tablenotes}[para,flushleft]
Note: We notice less training time and lower performance than that in Table \ref{table:Performance metrics and training time for Model 1 with time-distributed feature learning.}. The fall in performance is due to the lack of the time-distributed layer. We lack pseudo-temporal features that extract more information required in a successful classification. There is a significant drop in training time and performance for the CoS classifier, compared with the conventional one.
\end{tablenotes}
\label{table:Performance metrics and training time for Model 1 without time-distributed feature learning.}
\end{threeparttable}
\end{table}

Table \ref{table:Performance metrics and training time for Model 1 with time-distributed feature learning.} shows the performance metrics and training time for the CNN-TD(FFNN) for conventional NTC and CoS classification. We present the training time and performance metrics for the vanilla version of Model 1 in Table \ref{table:Performance metrics and training time for Model 1 without time-distributed feature learning.}. We notice that the performance of the new Model 1 is stable in all trials, resulting from the time-distribution wrapper. However, for the vanilla version, it is seen that the performance of the vanilla model highly depends on the initial hyper-parameters; these initial hyper-parameters are randomly set by the training process for every trial. We obtain the optimized hyper-parameters at the end of training, which will be used during the testing. Choosing initial random hyper-parameters is a common practice in deep learning \cite{probst2019tunability} to avoid bias. The convergence of time-distributed deep learning is robust, unlike the vanilla version. From Table \ref{table:Performance metrics and training time for Model 1 with time-distributed feature learning.}, the average \textit{F}1-scores are 88.5\% and 96.9\% for conventional and CoS classification, respectively. The average time duration to train Model 1 is 2.85 minutes, while for the vanilla model, it is 1.53 minutes for conventional classification. Model 1 averages 1.22 minutes, and the vanilla model utilizes 0.81 minutes for CoS classification. The observation shows that the vanilla version takes less time to train. The time complexity of deep learning depends on various factors such as model structure, forward propagation, backward propagation, the number of epochs in training, and data processing during training. Therefore, we provide a rough estimation of time complexity for Model 1, which can be given as $\mathcal{O}[128\times48\times(8\times6)\times(3\times3)]$, where 128 is the number of units, 48 is the number of features for Dataset 1, (8x6) size of modified data as explained in Section \ref{sec:DR}, (3x3) is kernel size of CNN. Based on the rough estimation, the time complexity for Model 1 is linear, which is easy and faster to execute. Also, the GPU enables faster convolution operation, which is the reason for less time to train the Model 1. The time-distribution version achieves stable and robust performance over the vanilla counterpart.

\subsubsection{Model 2}

\begin{table}
\begin{threeparttable}
\caption{Performance metrics and training time for Model 2.}
\centering 
\begin{tabular}
{
>{\centering\arraybackslash}m{0.9cm} 
>{\centering\arraybackslash}m{0.5cm} 
>{\centering\arraybackslash}m{0.8cm} >{\centering\arraybackslash}m{0.8cm} >{\centering\arraybackslash}m{0.8cm}
>{\centering\arraybackslash}m{0.8cm}
>{\centering\arraybackslash}m{0.8cm}
>{\centering\arraybackslash}m{0.5cm}
}
\hline
\multirow{7}{0.9cm}{\centering{Conven-\\tional}} &
Trial & Accuracy & Precision & Recall & F1 & Time (min) \\
\cline{2-7}
&1&0.906&0.906&0.906&0.898&32.55\\
&2& 0.883 & 0.880 & 0.883 & 0.873& 18.60\\
&3& 0.889 & 0.884  & 0.889 & 0.882 & 27.00\\
&4& 0.880 & 0.881  & 0.880 & 0.871 & 24.80\\
&\textbf{5}& \textbf{0.920} & \textbf{0.917}  & \textbf{0.920} & \textbf{0.912} & \textbf{31.00}\\
\cline{2-7}
&\textbf{Avg.} & \textbf{0.896} & \textbf{0.894} & \textbf{0.896} & \textbf{0.887} & \textbf{26.89}  \\
\hline
\multirow{7}{0.9cm}{\centering{CoS}} &
Trial & Accuracy & Precision & Recall & F1 & Time (min) \\
\cline{2-7}
&1        & 0.969  & 0.968  & 0.969  & 0.968                  & 13.35      \\
&2        & 0.959  & 0.957  & 0.959  & 0.956                  &\;\;8.80        \\
&3        & 0.965  & 0.965  & 0.965 & 0.963                 & 12.53      \\
&4        & 0.973  & 0.973  & 0.973  & 0.972                & 16.50       \\
&5        & 0.949  & 0.949  & 0.949  & 0.945                  & \;\;9.00          \\
\cline{2-7}
&\textbf{Avg.} & \textbf{0.963} & \textbf{0.962}   & \textbf{0.963} & \textbf{0.961}& \textbf{12.03}  \\
\hline
\end{tabular}
\begin{tablenotes}[para,flushleft]
Note: Model 2 takes more training time than Model 1 does because of internal components to find $\bm{\phi}$ of the LSTM, as shown in Eq. \eqref{Equ:mod2_E_7}. 
\end{tablenotes}
\label{table:Performance metrics and training time for Model 2 with time-distributed feature learning.}
\end{threeparttable}
\end{table}

\begin{table}
\begin{threeparttable}
\caption{Performance Metrics and Training Time for the Vanilla Version of Model 2 without Time-distributed Feature Learning.}
\centering 
\begin{tabular}
{>{\centering\arraybackslash}m{0.9cm} >{\centering\arraybackslash}m{0.5cm} >{\centering\arraybackslash}m{0.8cm} >{\centering\arraybackslash}m{0.8cm} >{\centering\arraybackslash}m{0.8cm}>{\centering\arraybackslash}m{0.8cm}>{\centering\arraybackslash}m{0.8cm}>{\centering\arraybackslash}m{0.5cm}}
\hline
\multirow{7}{0.9cm}{\centering{Conven-\\tional}} &
Trial & Accuracy & Precision & Recall & F1 & Time (min) \\
\cline{2-7}
&1      & 0.857  & 0.855  & 0.857 & 0.845                 & 24.65      \\
&2      & 0.830  & 0.829  & 0.830  & 0.815               & 16.68      \\
&3      & 0.766  & 0.761  & 0.766  & 0.757                  & \;\;7.83       \\
&4      & 0.817  & 0.816  & 0.817  & 0.811               & 14.25      \\
&\textbf{5}      & \textbf{0.885}  & \textbf{0.880}  & \textbf{0.885} & \textbf{0.873}          & \textbf{22.75}      \\
\cline{2-7}
&\textbf{Avg.}  & \textbf{0.831} & \textbf{0.828}& \textbf{0.831} & \textbf{0.820}             & \textbf{17.23}  \\
\hline
\multirow{7}{0.9cm}{\centering{CoS}} &
Trial & Accuracy & Precision & Recall & F1 & Time (min) \\
\cline{2-7}
&1      & 0.804 & 0.811 & 0.804 & 0.800                & \;\;9.50        \\
&2      & 0.854 & 0.853 & 0.854 & 0.852                & 10.61      \\
&3      & 0.845 & 0.843 & 0.845& 0.840                & 10.73      \\
&\textbf{4}      & \textbf{0.875} & \textbf{0.875}  & \textbf{0.875} & \textbf{0.875}                & \textbf{21.00}        \\
&5      & 0.863 & 0.861  & 0.863 & 0.860              & 19.28      \\
\cline{2-7}
&\textbf{Avg.} &\textbf{ 0.848} & \textbf{0.848}   & \textbf{0.848} & \textbf{0.845}               & \textbf{14.22} \\
\hline
\end{tabular}
\begin{tablenotes}[para,flushleft]
Note: As expected, the vanilla model without a time-distributed layer takes less time for training. However, the performance is not good compared with Model 2. 
\end{tablenotes}
\label{table:Performance metrics and training time for Model 2 without time-distributed feature learning.}
\end{threeparttable}
\end{table}

Table \ref{table:Performance metrics and training time for Model 2 with time-distributed feature learning.} shows the performance metrics and training time for Model 2 (LSTM-TD(FFNN)) for both conventional and CoS NTC. Table \ref{table:Performance metrics and training time for Model 2 without time-distributed feature learning.} shows the performance and training time of the vanilla version, i.e., LSTM-FFNN. From Table \ref{table:Performance metrics and training time for Model 2 with time-distributed feature learning.}, we can notice that the \textit{F}1-scores of Model 2 are 88.7\% and 96.1\% for conventional and CoS classification, respectively, higher than those of the vanilla model given by Table \ref{table:Performance metrics and training time for Model 2 without time-distributed feature learning.}. The average time duration to train the model is 26.89 minutes for conventional classification, while for the vanilla model without time-distributed learning, it is 17.23 minutes. For the CoS classification, Model 2 spends an average training time of 12 minutes, comparable to the 14.22 minutes spent by the vanilla version. Compared with the metrics of Model 1 given by Table \ref{table:Performance metrics and training time for Model 1 with time-distributed feature learning.}, the performance of Model 2 is similar. However, the LSTM-based Model 2 takes longer time for training since there is the cost for finding the function $\bm{\phi}$. However, the time-distribution methodology is certainly enhancing the performance and stability of NTCs.


\subsubsection{Model 3}

\begin{table}
\begin{threeparttable}
\caption{Performance Metrics and Training Time for Model 3.}
\centering 
\begin{tabular}
{
>{\centering\arraybackslash}m{0.9cm} 
>{\centering\arraybackslash}m{0.5cm} 
>{\centering\arraybackslash}m{0.8cm} >{\centering\arraybackslash}m{0.8cm} >{\centering\arraybackslash}m{0.8cm}
>{\centering\arraybackslash}m{0.8cm}
>{\centering\arraybackslash}m{0.8cm}
>{\centering\arraybackslash}m{0.5cm}
}
\hline
\multirow{7}{0.9cm}{\centering{Conven-\\tional}} &
Trial & Accuracy & Precision & Recall & F1 & Time (min) \\
\cline{2-7}
&1      & 0.937 & 0.935 & 0.937 & 0.934               & 17.15      \\
&\textbf{2}      & \textbf{0.944} & \textbf{0.942} & \textbf{0.944} & \textbf{0.941}               & \textbf{21.20}       \\
&3      & 0.938 & 0.936  & 0.938 & 0.936               & 17.46      \\
&4      & 0.941 & 0.939  & 0.941 & 0.939               & 21.48      \\
&5      & 0.939 & 0.937  & 0.939 & 0.937                & 20.96      \\
\cline{2-7}
&\textbf{Avg.} & \textbf{0.940} & \textbf{0.938}& \textbf{0.940} & \textbf{0.937}             & \textbf{19.65}     \\
\hline
\multirow{7}{0.9cm}{\centering{CoS}} &
Trial & Accuracy & Precision & Recall & F1 & Time (min) \\
\cline{2-7}
&\textbf{1}      & \textbf{0.994} & \textbf{0.994} & \textbf{0.994} & \textbf{0.994}                & \textbf{13.53}      \\
&2      & 0.993 & 0.993 & 0.993& 0.993                & 10.61      \\
&3      & 0.993 & 0.992  & 0.993 & 0.992                 & \;\;6.50        \\
&4      & 0.993 & 0.993 & 0.993 & 0.993                 & \;\;7.80        \\
&5      & 0.993 & 0.993  & 0.993 & 0.993                & \;\;8.20       \\
\cline{2-7}
&\textbf{Avg.} & \textbf{0.996} & \textbf{0.993} & \textbf{0.996} & \textbf{0.993}              & \textbf{\;\;9.32} \\
\hline
\end{tabular}
\begin{tablenotes}[para,flushleft]
Note: We can notice a significant increase of 5\% in the performance. The average training time is 19.65 min. It is evident that the extraction of holistic-temporal features helps improve performance.
\end{tablenotes}
\label{table:Performance metrics and training time for Model 3 with time-distributed feature learning.}
\end{threeparttable}
\end{table}

\begin{table}
\begin{threeparttable}
\caption{Performance metrics and training time for Model 3 without time-distributed feature learning.}
\centering 
\begin{tabular}
{
>{\centering\arraybackslash}m{0.9cm} 
>{\centering\arraybackslash}m{0.5cm} 
>{\centering\arraybackslash}m{0.8cm} >{\centering\arraybackslash}m{0.8cm} >{\centering\arraybackslash}m{0.8cm}
>{\centering\arraybackslash}m{0.8cm}
>{\centering\arraybackslash}m{0.8cm}
>{\centering\arraybackslash}m{0.5cm}
}
\hline
\multirow{7}{0.9cm}{\centering{Conven-\\tional}} &
Trial & Accuracy & Precision & Recall & F1 & Time (min) \\
\cline{2-7}
&1      & 0.834 & 0.830  & 0.834 & 0.823               & 14.11      \\
&2      & 0.799 & 0.796  & 0.799 & 0.792               & 12.50       \\
&3      & 0.851 & 0.851  & 0.851 & 0.838                & 14.33      \\
&\textbf{4}      & \textbf{0.890} & \textbf{0.886}  & \textbf{0.890} & \textbf{0.882}               & \textbf{14.55}      \\
&5      & 0.885 & 0.884  & 0.885 & 0.873               & 14.55      \\
\cline{2-7}
&\textbf{Avg.} & \textbf{0.852} & \textbf{0.849} & \textbf{0.852} & \textbf{0.841}              & \textbf{14.00} \\
\hline
\multirow{7}{0.9cm}{\centering{CoS}} &
Trial & Accuracy & Precision & Recall & F1 & Time (min) \\
\cline{2-7}
&1      & 0.862 & 0.897 & 0.862 & 0.854                & 5.13       \\
&2      & 0.830 & 0.848  & 0.830 & 0.826               & 4.21       \\
&\textbf{3}      & \textbf{0.921} & \textbf{0.923} & \textbf{0.921} & \textbf{0.919}                & \textbf{5.41}      \\
&4      & 0.763 & 0.822  & 0.763 & 0.769                & 5.25       \\
&5      & 0.884 & 0.903  & 0.884 & 0.883               & 4.60        \\
\cline{2-7}
&\textbf{Avg.} & \textbf{0.852} & \textbf{0.878}  & \textbf{0.852} & \textbf{0.850}             &\textbf{4.92}    \\
\hline
\end{tabular}
\begin{tablenotes}[para,flushleft]
Note: The performance of the vanilla counterpart of Model 3 without time-distributed learning is significantly lower for conventional NTC. The performance for CoS classification without time-distributed learning is lower. Training time is shorter than Model 3.
\end{tablenotes}
\label{table:Performance metrics and training time for Model 3 without time-distributed feature learning.}
\end{threeparttable}
\end{table}

Table \ref{table:Performance metrics and training time for Model 3 with time-distributed feature learning.} shows the performance metrics and duration of training for Model 3 or CNN-LSTM-TD(FFNN) for conventional and CoS NTCs. Table \ref{table:Performance metrics and training time for Model 3 without time-distributed feature learning.} shows those for the vanilla model or CNN-LSTM-FFNN. From Table \ref{table:Performance metrics and training time for Model 3 with time-distributed feature learning.}, the \textit{F}1-scores are 93.7\%  and 99.3\% for conventional and CoS classification. The average time to train the Model 3 is 19.65 minutes for conventional classification, and the vanilla model is 14.22 minutes. For CoS classification, Model 3 spends an average time of 9.32 minutes, compared with an average of 4.92 minutes for the vanilla model.

As expected, the performance of Model 3 or CNN-LSTM-TD(FFNN) is the best among all the models. Model 3 shows overall better performance in both conventional and CoS classification. Comparing the training time with other models, we notice that Model 3 takes a shorter time than Model 2. The reason is that the initial hyper-parameter values and data dimension cause the LSTM $\bm{\phi}$ approximation in Model 2 to be slower. Moreover, in Model 3, the CNN down-converts the data for LSTM, resulting in a shorter time in training. 

\subsection{Classification report per class for Model 3 for Multi-class NTC}

\begin{figure*}
 	\centering
 	\includegraphics[width=\linewidth]{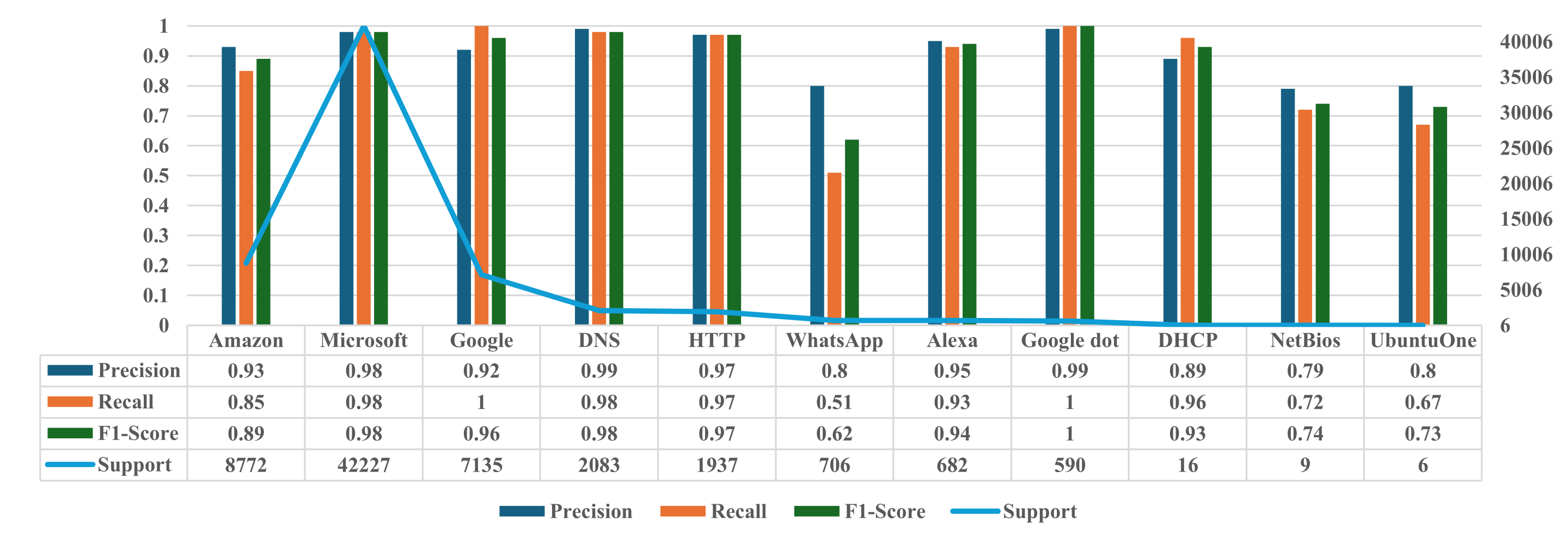} 
 	\caption{Multi-class classification report for Model 3 using Dataset 1. Support in the figure represents the number of test samples.}
 	\label{fig:mc}
\end{figure*}

We study the multi-class classification report for the conventional NTC problem to decipher the performance of our solution for various application protocols that are classified in the problem. Dataset 1 consists of 141 classes; however, in the report, we consider a few protocols with more, medium, and less number of test samples in the experiment. Figure \ref{fig:mc} shows the report with several samples at the bottom as support. We provide precision, recall, and F1 scores for each application protocol used in the work. For WhatsApp, the F1-score is low; however, Microsoft and other application protocols are identified well with good F1-scores.

\color{black}

\subsection{Generalization Performance of the solution on different datasets}
\label{othdataset}

\begin{figure}
 	\centering
 	\includegraphics[width=\linewidth]{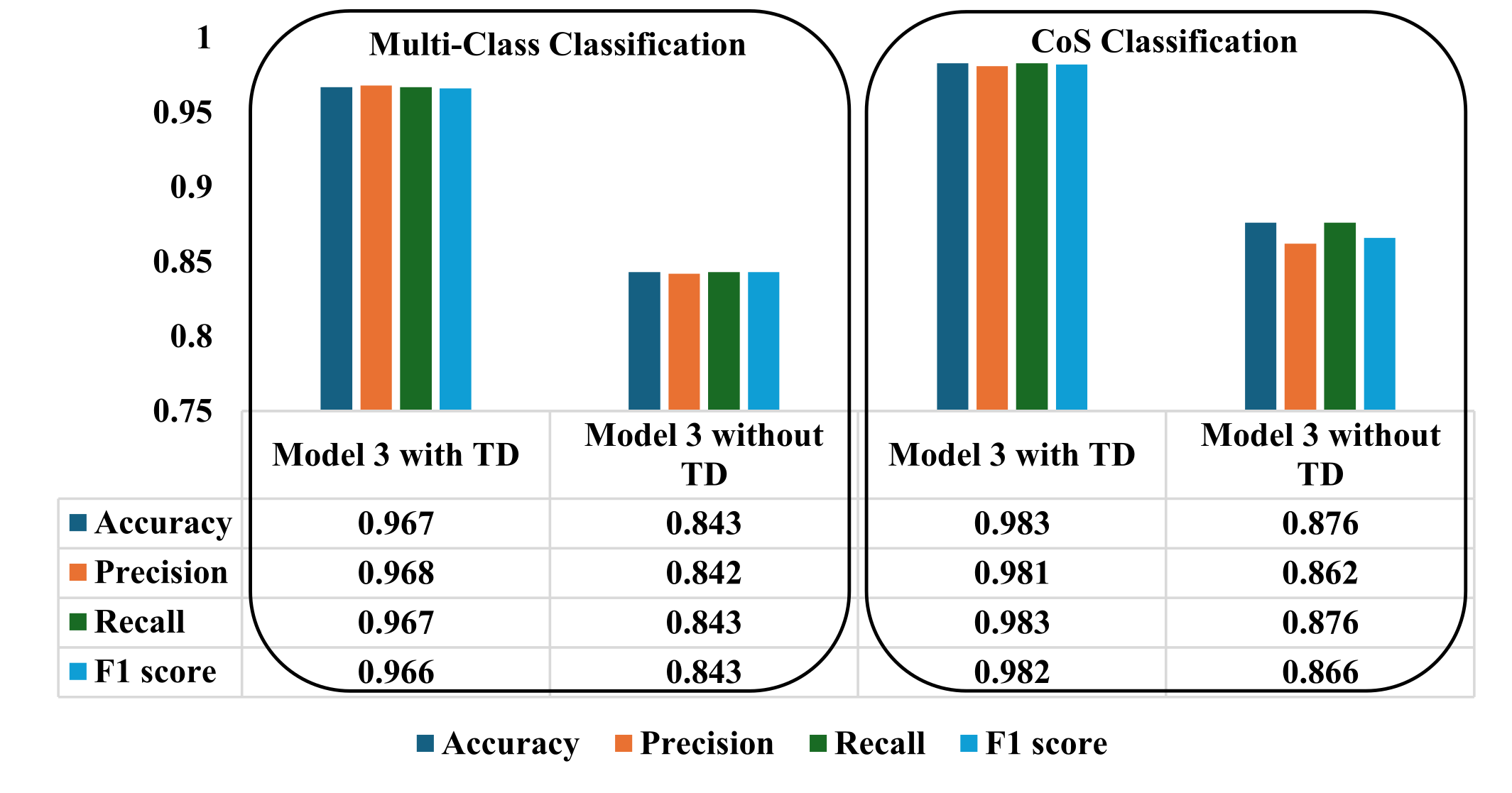} 
 	\caption{Performance Metrics of TD and non-TD versions of Model 3 for Dataset 2 for multi-class and class-of-service network traffic classifications.}
 	\label{fig:dt2}
\end{figure}

In this section, we study the generalization of the solution on different datasets. We provide the average results from five training and testing trials for the rest of the datasets for Model 3. Model 3 is the solution that extracts holistic-temporal features. Figure \ref{fig:dt2} shows the performance metrics for Dataset 2 for TD and non-TD versions of Model 3 for multi-class network traffic classification. The TD version of Model 3 is 14\% better than the non-TD version for multi-class classification. We can notice a 12\% increase in the performance of the TD version of Model 3 for CoS classification.

\begin{figure}
 	\centering
 	\includegraphics[width=\linewidth]{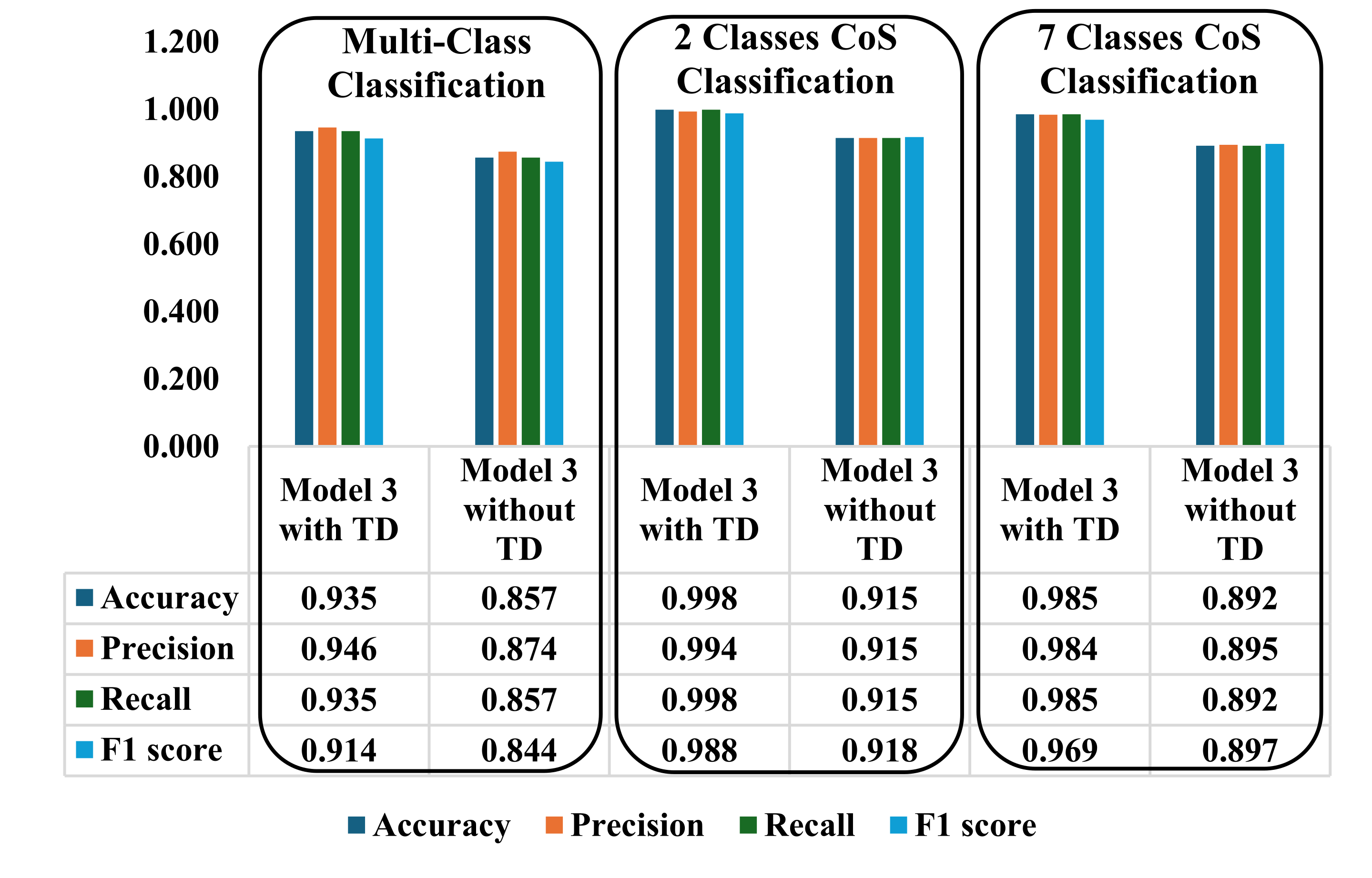} 
 	\caption{Performance Metrics of TD and non-TD versions of Model 3 for Dataset 3 for multi-class and CoS network traffic classifications. Dataset 3 consists of binary and seven classes of CoS problems. Please refer to Table \ref{table:dataset}.}
 	\label{fig:dt3}
\end{figure}

Figure \ref{fig:dt3} gives the performance metrics for Dataset 3 for TD and non-TD versions of Model 3 for multi-class and two classes as well as seven classes of CoS network traffic classifications. We can notice a 9\% improvement in the TD version's performance over the non-TD version of Model 3 for multi-class NTC and for binary CoS classifications on Dataset 3. The TD version of Model 3 shows a 10\% improvement in performance for seven classes CoS NTC. 

\begin{figure}
 	\centering
 	\includegraphics[width=\linewidth]{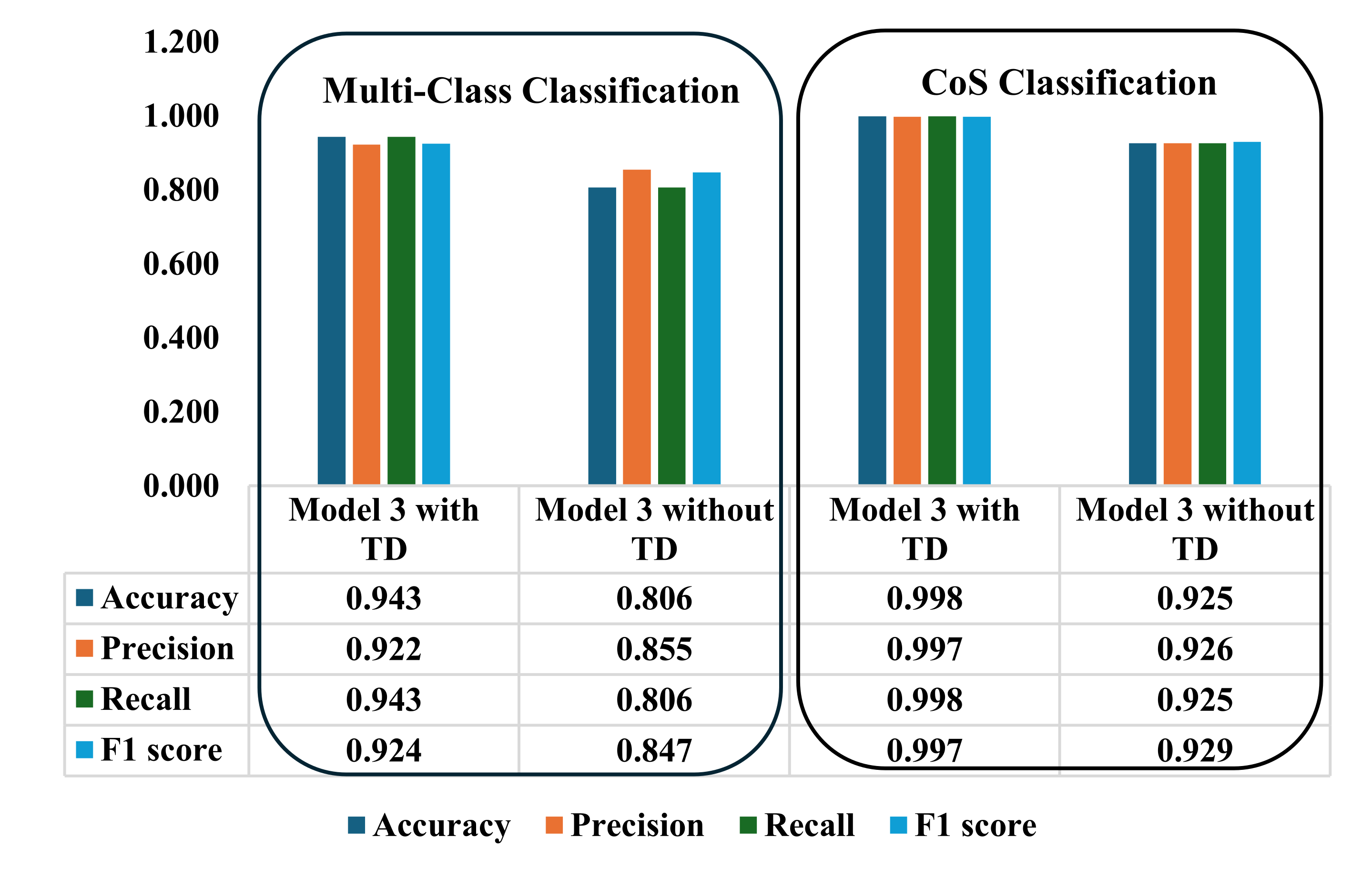} 
 	\caption{Performance Metrics of TD and non-TD versions of Model 3 for Dataset 4 for multi-class and CoS network traffic classifications.}
 	\label{fig:dt4}
\end{figure}

The performance metrics of Model 3 with TD and non-TD versions for Dataset 4 are given in Figure \ref{fig:dt4}. 16\% improvement is marked by the TD version of Model 3 over the non-TD version of Model 3 for multi-class classification. The TD version of Model 3 shows a 7\% improvement in performance for CoS NTC.

\color{black}

\subsection{Cluster visualization}
\begin{figure}
 	\centering
 	\includegraphics[trim={2 2 2 2},width=\linewidth]{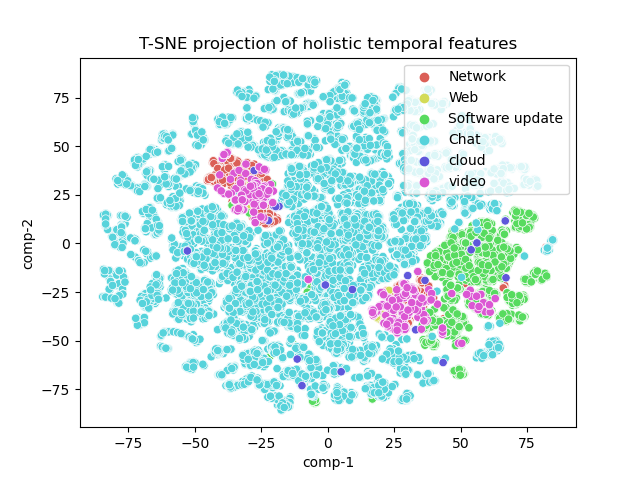} 
 	\caption{T-SNE projection of holistic-temporal features from the new Model 3. We find the classes are more tightly clustered with fewer outliers, resulting in better classification.}
 	\label{fig:tsne1}
\end{figure}

\begin{figure}
 	\centering
 	\includegraphics[trim={2 2 2 2},width=\linewidth]{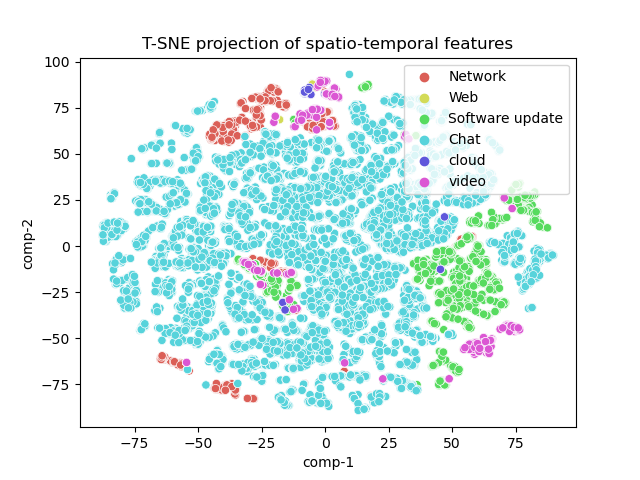} 
 	\caption{T-SNE projection of spatio-temporal features from the vanilla version of Model 3 with no time-distributed layer. Spatio-temporal features also provide tighter clusters. However, two or more classes are merged, indicating poor performance.}
 	\label{fig:tsne2}
\end{figure}

Furthermore, to understand the degree of data separability, we use the T-SNE \cite{van2008visualizing} reduction technique to visualize the features before the Decision-making layer. All features will be reduced to two components of T-SNE as explained in \cite{van2008visualizing}. If the T-SNE depicts lots of overlaps in clusters of classes, then the classification results will be less accurate. On the contrary, well-separated data clusters on the T-SNE graph provide better results. Therefore, we plot the T-SNE graphs for the features extracted by the new Model 3 to visualize the fruition of the holistic-temporal features in Fig. \ref{fig:tsne1}, and the projection of spatio-temporal features that are extracted from the vanilla model with no time-distribution layer is shown in Fig. \ref{fig:tsne2}. We plot the graphs for CoS classification for six unbalanced classes. Due to hardware limitations, not all the classes are projected on the graphs. From the figures, we can notice the overlapping among the data clusters in the spatio-temporal features of Fig. \ref{fig:tsne2} is more than the holistic-temporal features in Fig. \ref{fig:tsne1}. Every class is aggregated well in Fig. \ref{fig:tsne1}, providing better results. On the contrary, classes are scattered with more overlaps in Fig. \ref{fig:tsne2}. The two figures show the superior classification performance of the new Model 3. The observation is aligned with the results presented earlier.

\subsection{Performance Comparison and Discussion} 

As mentioned earlier, we consider the works \cite{lopez2017network} and \cite{d2021network} as the SOA solutions for conventional and CoS classification, respectively. Comparing the works in NTC is challenging mainly because of the differences in the data. Therefore, we implement the model and evaluate the models using the same dataset. We implement the two SOA models from \cite{lopez2017network} and \cite{d2021network} and denote them as SOA 1 and SOA 2, respectively. Then, we use our Dataset 1 to evaluate the models. We use 70\% of the dataset as the training set, 10\% for validation, and the rest 20\% for testing the models. The two SOA models are tested for both the conventional problem and the CoS classification problem. We focus on comparing Model 3 with the two SOA models since it performs best among the three models, as shown in Section \ref{sec:comparisonvanilla}.

\begin{figure}
 	\centering
 	\includegraphics[width=0.99\linewidth]{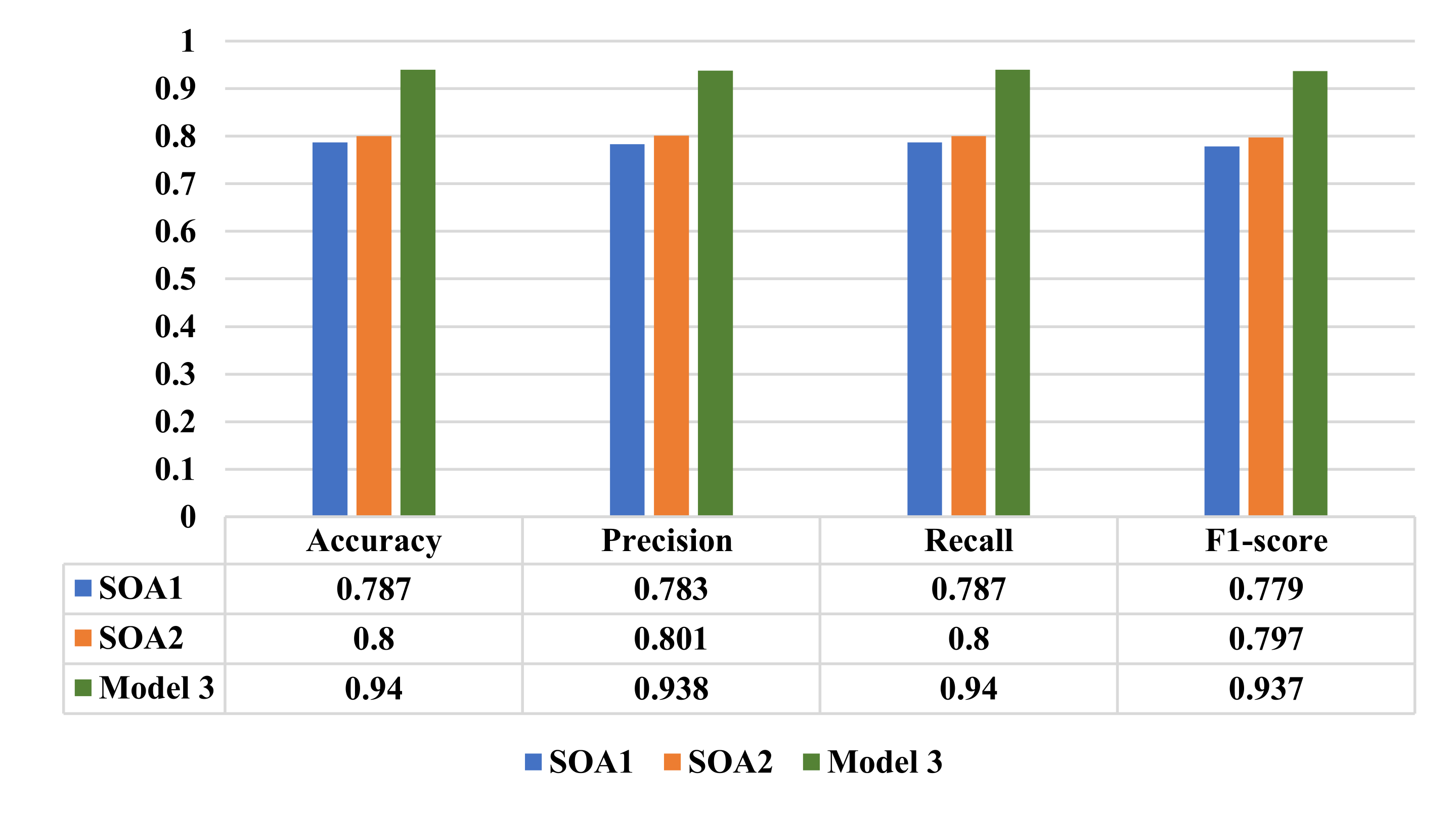} 
 	\caption{Performance comparison of the SOA models with the proposed new time-distributed model (Model 3) for conventional classification. Our solution achieves the best performance with an accuracy of 94\%, while SOA1 and SOA2 achieves 78.7\% and 89\%, respectively.}
 	\label{fig:conventional comparision}
\end{figure}

\begin{figure}
 	\centering
 	\includegraphics[width=0.99\linewidth]{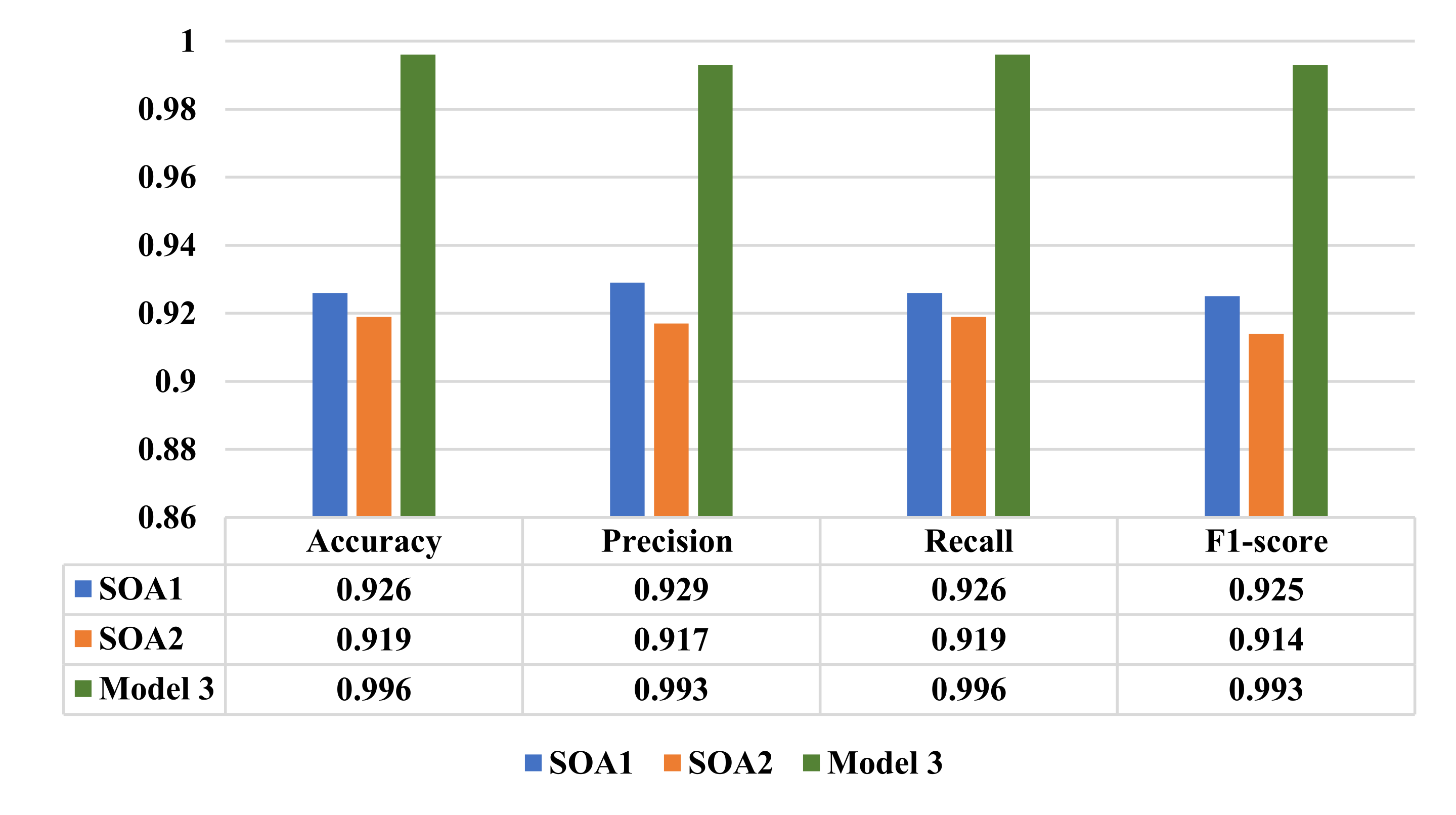} 
 	\caption{Performance comparison of SOA models with new Model 3 for CoS classification. Our solution shows exceptional performance for CoS classification with an accuracy of 99.6\%. SOA1 is 92.6\% accurate and SOA2 shows 91.9\% accuracy for the CoS classification problem.}
 	\label{fig:class-of-service comparision}
\end{figure}

Fig. \ref{fig:conventional comparision} depicts the comparison for conventional traffic classification. We show the average values of performance metrics in the figure. Our new Model 3 shows improvement in \textit{F}-1 score of 20\% better than SOA 1 and 17\% better than SOA 2. Fig. \ref{fig:class-of-service comparision} compares the average performance metrics of SOA models with our model for CoS classification. Our model is 7\% better than the SOA 1 model and 8\% better than the SOA 2 model in terms of \textit{F}1 score.

 
 Table \ref{table:Performance metrics and training time for conventional NTC for SOA 1 and SOA 2} shows the performance metrics and training time of SOA 1 for conventional and CoS NTCs. Table \ref{table:Performance metrics and training time for CoS NTC for SOA 1 and SOA 2} shows the performance metrics and training time for conventional and CoS NTCs for SOA 2. We notice that the new time-distributed model is more stable in all trials as given by Table \ref{table:Performance metrics and training time for Model 3 with time-distributed feature learning.} in comparison with SOA 1 and SOA 2. On the other hand, the SOA model's performances vary in different trials as shown by Tables \ref{table:Performance metrics and training time for conventional NTC for SOA 1 and SOA 2} and \ref{table:Performance metrics and training time for CoS NTC for SOA 1 and SOA 2}. Therefore, our method is more robust and resistant to the initial hyper-parameters. 
Regarding the performance metrics, our model outperforms both models in conventional and CoS classification problems. For conventional NTC, Model 3 shows a 19.44\% improvement in accuracy compared with SOA 1, and 7.55\% more accurate than SOA 2. On average, our solution is 13.5\% more accurate than the SOA models. Although the total time taken for training by the SOA models is shorter than the new model, it is a worthwhile trade-off, and the new model's performance is significantly better.

\begin{table}
\begin{threeparttable}
\caption{Performance metrics and training time for SOA 1}
\centering 
\begin{tabular}
{
>{\centering\arraybackslash}m{0.9cm} 
>{\centering\arraybackslash}m{0.5cm} 
>{\centering\arraybackslash}m{0.8cm} >{\centering\arraybackslash}m{0.8cm} >{\centering\arraybackslash}m{0.8cm}
>{\centering\arraybackslash}m{0.8cm}
>{\centering\arraybackslash}m{0.8cm}
>{\centering\arraybackslash}m{0.5cm}
}
\hline
\multirow{7}{0.9cm}{\centering{Conven-\\tional}} &
Trial & Accuracy & Precision & Recall & F1 & Time (min) \\
\cline{2-7}
&\textbf{1}      & \textbf{0.875}  & \textbf{0.870}  & \textbf{0.875}  & \textbf{0.869}                & \textbf{17.15}      \\
&2      & 0.862  & 0.860  & 0.862  & 0.854                 & \;\;4.08       \\
&3      & 0.627  & 0.616  & 0.627 & 0.615                & 13.50       \\
&4      & 0.745  & 0.744  & 0.745  & 0.736                 & 14.13      \\
&5      & 0.827  & 0.825  & 0.827  & 0.819                & 15.10       \\
\cline{2-7}
&\textbf{Avg.} & \textbf{0.787} & \textbf{0.783} & \textbf{0.787} & \textbf{0.779}              & \textbf{12.79}  \\
\hline
\multirow{7}{0.9cm}{\centering{CoS}} &
Trial & Accuracy & Precision & Recall & F1 & Time (min) \\
\cline{2-7}

&1      & 0.929 & 0.928  & 0.929 & 0.928                & 4.40        \\
&2      & 0.933 & 0.933  & 0.933 & 0.932                & 4.60        \\
&3      & 0.885 & 0.897  & 0.885 & 0.884                & 5.10        \\
&4      & 0.937 & 0.939  & 0.937 & 0.937               & 4.70        \\
&\textbf{5}      & \textbf{0.946} & \textbf{0.945}  & \textbf{0.946} & \textbf{0.945}                & \textbf{5.15}       \\
\cline{2-7}
&\textbf{Avg.} & \textbf{0.926} & \textbf{0.929} & \textbf{0.926} & \textbf{0.925}             & \textbf{4.79} \\

\hline
\end{tabular}
\begin{tablenotes}[para,flushleft]
\end{tablenotes}
\label{table:Performance metrics and training time for conventional NTC for SOA 1 and SOA 2}
\end{threeparttable}
\end{table}

\begin{table}
\begin{threeparttable}
\caption{Performance metrics and training time for SOA 2}
\centering 
\begin{tabular}
{
>{\centering\arraybackslash}m{0.9cm} 
>{\centering\arraybackslash}m{0.5cm} 
>{\centering\arraybackslash}m{0.8cm} >{\centering\arraybackslash}m{0.8cm} >{\centering\arraybackslash}m{0.8cm}
>{\centering\arraybackslash}m{0.8cm}
>{\centering\arraybackslash}m{0.8cm}
>{\centering\arraybackslash}m{0.5cm}
}
\hline
\multirow{7}{0.9cm}{\centering{Conven-\\tional}} &
Trial & Accuracy & Precision & Recall & F1 & Time (min) \\
\cline{2-7}

&1      & 0.789 & 0.789  & 0.789 & 0.774               & 15.33      \\
&2      & 0.744 & 0.731  & 0.744 & 0.752               & 14.00         \\
&3      & 0.739 & 0.731  & 0.739 & 0.727              & 14.30       \\
&\textbf{4}     & \textbf{0.870} & \textbf{0.875} & \textbf{0.870} & \textbf{0.857}               & \textbf{16.00}         \\
&5      & 0.858 & 0.878  & 0.858 & 0.875                & 15.90       \\
\cline{2-7}
&\textbf{Avg.}  & \textbf{0.800} & \textbf{0.801}  & \textbf{0.800} & \textbf{0.797}                & \textbf{15.11} \\
\hline
\multirow{7}{0.9cm}{\centering{CoS}} &
Trial & Accuracy & Precision & Recall & F1 & Time (min) \\
\cline{2-7}

&\textbf{1}      & \textbf{0.959} & \textbf{0.956}  & \textbf{0.959} & \textbf{0.957}                & \textbf{9.06}       \\
&2      & 0.907 & 0.911  & 0.907 & 0.906                & 8.25       \\
&3      & 0.935 & 0.931  & 0.935 & 0.932              & 8.80        \\
&4      & 0.944 & 0.944  & 0.944 & 0.943                & 9.90        \\
&5      & 0.850 & 0.845  & 0.850 & 0.834                & 8.10        \\
\cline{2-7}
&\textbf{Avg.}  & \textbf{0.919} & \textbf{0.917}  & \textbf{0.919} & \textbf{0.914}                & \textbf{8.82} \\
\hline
\end{tabular}
\begin{tablenotes}[para,flushleft]
\end{tablenotes}
\label{table:Performance metrics and training time for CoS NTC for SOA 1 and SOA 2}
\end{threeparttable}
\end{table}

\subsection{Analysis of Trainable Parameters}

The computation load of feature extraction is directly proportional to the number of parameters required to learn during training. We established that the time-distributed feature learning method introduces more features in the classification tasks; therefore, it is crucial to analyze the extra parameters to assess the cost of training. Section \ref{sec:math_model} introduces a mathematical explanation of the extra features derived in the classification tasks. However, we present the quantification of the number of trainable parameters in this section. 

The number of learnable parameters in the CNN is given by $(P \times Q \times S + B) \times U$, where $P$ and $Q$ are the vertical and horizontal sizes of the selected convolutional kernel, respectively, $S$ is the input size, $B$ is the bias size, and $U$ is the number of units. Max-pooling down-samples the inputs, so there are no parameters to learn. The gamma and beta weights are learnable parameters in batch normalization; therefore, $2\times U$ parameters are required to learn at batch normalization. The number of trainable parameters in the LSTM is $4 \times \left[ (S+1) \times U + U^2\right]$. $S \times U + U$ parameters must be trained for an FFNN.


\begin{table}
\begin{threeparttable}

\caption{Number of trainable parameters for Model 3.}
\centering 
\begin{tabular}{
>{\centering\arraybackslash}m{1.5cm} 
>{\centering\arraybackslash}m{3.9cm} >{\centering\arraybackslash}m{1.6cm}} 
\hline
Network & Calculation & Trainable parameters \\
\hline 
CNN\_2D & $(3\times3\times1+1)\times128$ & 1280\\
MP\_2D\ & - & 0 \\
BN & $2\times128$ &  256\\
Reshape & - & 0\\
LSTM & $4\times\left[(128+128)\times128+128^2\right]$ & 131584\\
TD(FFNN\_0) & $128\times128+128$ & 16512\\
Flatten & - & 0 \\
FFNN\_1 & $6\times128\times141+141$ & 108429 \\

\hline
\textbf{Total} & &\textbf{258,061}\\
\hline
\end{tabular}

\begin{tablenotes}[para,flushleft]
\end{tablenotes}
\label{table:Calculation of parameters of TD model 3}
\end{threeparttable}
\end{table} 

\begin{table}
\begin{threeparttable}

\caption{Number of Trainable Parameters for The Vanilla Version of Model 3.}
\centering 
\begin{tabular}{
>{\centering\arraybackslash}m{1.5cm} 
>{\centering\arraybackslash}m{3.9cm} >{\centering\arraybackslash}m{1.6cm}} 
\hline
Network & Calculation & Trainable parameters \\
\hline 
CNN\_2D & $(3\times3\times1+1)\times128$ & 1280\\
MP\_2D\ & - & 0 \\
BN & $2\times128$ &  256\\
Reshape & - & 0\\
LSTM & $4\times\left[(128+128)\times128+128^2\right]$ & 131584\\
FFNN\_0 & $128\times128+128$ & 16512\\
Flatten & - & 0 \\
FFNN\_1 & $128\times141+141$ & 18189 \\

\hline
\textbf{Total} & &\textbf{167,821}\\
\hline
\end{tabular}

\begin{tablenotes}[para,flushleft]
\end{tablenotes}
\label{table:Calculation of parameters of model 3}
\end{threeparttable}
\end{table}

We investigate the number of trainable parameters of Model 3 because of its excellent performance and the overhead in the training time. Table \ref{table:Calculation of parameters of TD model 3} shows the number of trainable parameters at every layer in Model 3. We notice that $6 \times 128$ extra parameters are added because of the time-distributed layer in the previous FFNN,  as is explained mathematically in Section \ref{sec:model3}. On the other hand, from Table \ref{table:Calculation of parameters of model 3}, which shows the trainable parameters for the vanilla version, we notice that those extra parameters are not present. In summary, Model 3 requires $258,061$ parameters to be trained, and its vanilla counterpart model needs $167,821$ parameters to be trained. $42\%$ percent of the parameters are added in Model 3 compared with its vanilla version without time-distributed learning. With more parameters, more features can be learned, resulting in better performance. Features extracted at the final stage are complex to explain; hence, we name them pseudo-temporal features.



\section{Conclusion and Future works}
\label{sec:conclusion}

Using time-distributed feature learning, we propose a novel, stable, and robust methodology for IoT traffic classification tasks and in general to IP network traffic classification. Our work shows that traffic data exhibit pseudo-temporal characteristics along with spatial-temporal features. By exploiting the current deep-learning models using a time-distributed wrapper, we can enhance the performance of traffic classifiers by leaps and bounds. Our solution not only stands as a superior one but also provides a general solution for multi-class and CoS types of classification. We explain the fruition of our method theoretically and experimentally using challenging real-world datasets of various types. 

To summarize, the newly proposed model, CNN-LSTM-TD(FFNN), outperforms the state-of-the-art performance for both types of NTCs with phenomenal results. Specifically, it achieves 94\%, which is 20\% better than state-of-the-art solutions for conventional traffic classification, and 99\% accuracy, which is 8\% better than the state-of-the-art solutions for CoS classification. On average, our solution is 13.5\% better than the current solutions. 

In the future, we need to test the solution in a live IoT network to determine the number of traffic samples required from every class for successful classification. Moreover, we must find a lightweight method to enable time-distributed feature learning. We still need to determine efficient features that do not require flow processing software to avoid features like IPs. Given the success of time-distributed feature learning in network traffic classification, we aim to explore the solution's adaptability in classifying time-series data from other domains.

\bibliographystyle{IEEEtran}
\bibliography{IEEEabrv,refs}
%

\vspace{-4.5cm}
\begin{IEEEbiography}
    [{\includegraphics[width=1in,height=1.25in,clip,keepaspectratio]{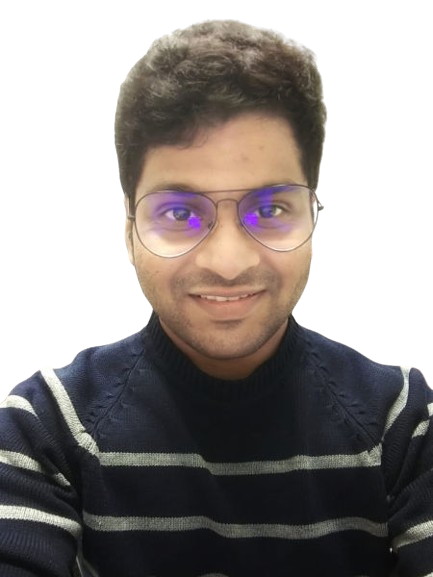}}]{Yoga Suhas Kuruba Manjunath}
    (Graduate student Member, IEEE) is currently a PhD candidate at Toronto Metropolitan University (TMU), formerly Ryerson University, Canada. He holds a Bachelor of Engineering in Electronics and Communication from Visvesvaraya Technological University (2014) and a Master of Engineering with a specialization in Artificial Intelligence (AI) from TMU (2021). With over five years of industry experience as an Internet-of-Things (IoT) architect, Yoga has developed IoT stacks for the dairy and hospitality industries, impacting over two hundred thousand customers. His unique blend of hardware and software expertise has been pivotal in numerous projects within these sectors.

    Upon returning to academia, Yoga has focused on advancing his research skills, resulting in publications in leading conferences and journals, including IEEE GLOBECOM, IEEE WF-IOT, and Electronic Commerce Research and Applications. He also contributes as a peer reviewer for several prestigious journals, such as IEEE Transactions on Wireless Communication, Electronic Commerce Research and Applications, and the Journal of Supercomputing. Yoga's active involvement in IEEE underscores his strong professional network and community engagement. He serves as Secretary of the IEEE Vehicular Technology Chapter at the IEEE Toronto section and is an active member of the IEEE Vehicular Technology Society (VTS) and IEEE Communication Society (ComSoc). Among his academic accolades is the "Best Team Award" at IEEE's Leaders of Tomorrow event, organized by IEEE Toronto.

    Yoga's current research interests include AI-based IoT solutions and Virtual Reality network optimization for Quality of Service (QoS). He actively participates in communications-related projects at the Communications and Signal Processing Applications Laboratory (CASPAL) and Ubiquitous Intelligent Communication and Computing (UICC) at TMU.

\end{IEEEbiography}

\vspace{-4.5cm}

\begin{IEEEbiography}
    [{\includegraphics[width=1in,height=1.25in,clip,keepaspectratio]{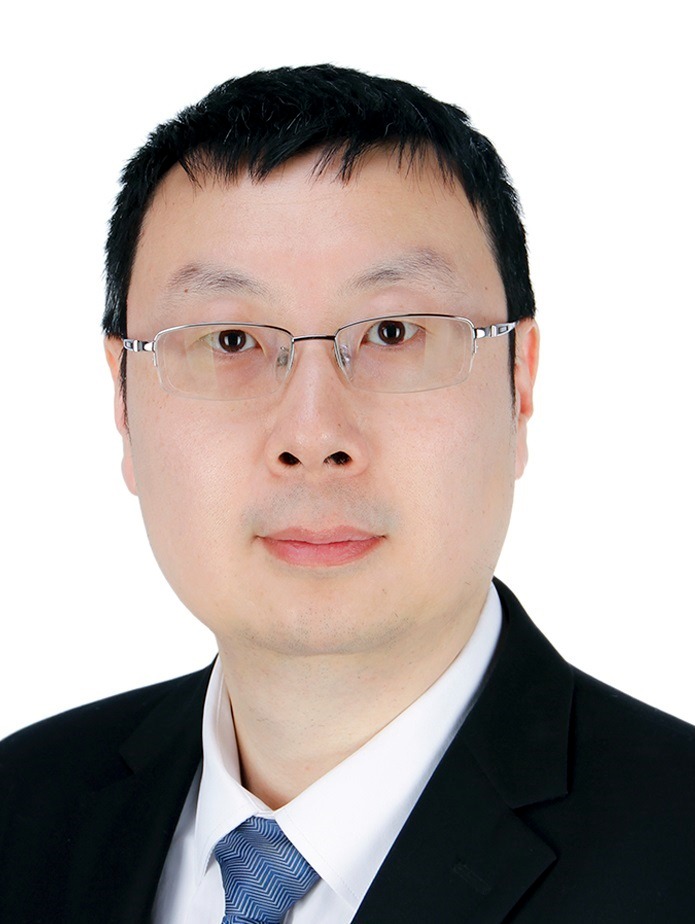}}]{Sihao Zhao}
    (Senior Member, IEEE) received the B.S. and Ph.D. degrees in Electronic Engineering from Tsinghua University, Beijing, China, in 2005 and 2011, respectively. He is a registered Professional Engineer in Ontario. He is currently a Senior Algorithm Designer with NovAtel, Autonomy and Positioning division of Hexagon, Calgary, AB, Canada. From 2011 to 2013, he was an Electronics Systems Engineer with China Academy of Space Technology, Beijing, China. From 2013 to 2019, he was a Postdoctoral Researcher and then an Assistant Professor with the Department of Electronic Engineering, Tsinghua University, Beijing, China. From 2020 to 2021, he was a Research Associate with the Department of Electrical, Computer and Biomedical Engineering, Toronto Metropolitan University, Toronto, ON, Canada. His research interests include localization algorithms, high-precision positioning techniques, and indoor navigation system development. He is also an Editor of GPS Solutions.

\end{IEEEbiography}

\begin{IEEEbiography}
    [{\includegraphics[width=1in,height=1.25in,clip,keepaspectratio]{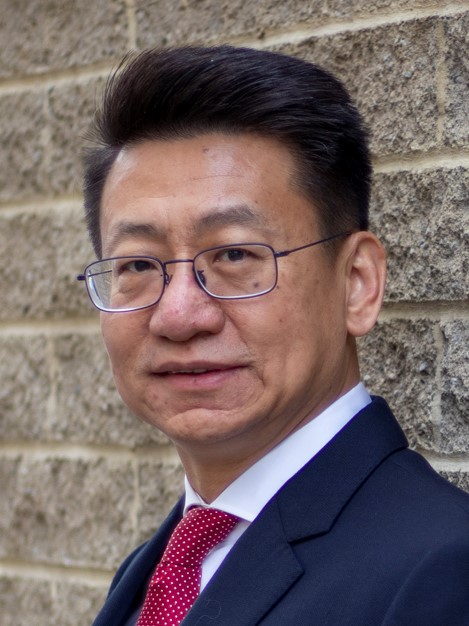}}]{Xiao-Ping Zhang}
    (Fellow, IEEE) received B.S. and Ph.D. degrees from Tsinghua University, in 1992 and 1996, respectively, both in Electronic Engineering. He holds an MBA in Finance, Economics and Entrepreneurship with Honors from the University of Chicago Booth School of Business, Chicago, IL.
    
    He is Chair Professor at Tsinghua Shenzhen International Graduate School (SIGS) and Tsinghua-Berkeley Shenzhen Institute (TBSI), Tsinghua University. He was the founding Dean of Institute of Data and Information (iDI) at Tsinghua SIGS. He had been with the Department of Electrical, Computer and Biomedical Engineering, Toronto Metropolitan University (Formerly Ryerson University), Toronto, ON, Canada, as a Professor and the Director of the Communication and Signal Processing Applications Laboratory (CASPAL), and has served as the Program Director of Graduate Studies. His research interests include image and multimedia content analysis, sensor networks and IoT, machine learning/AI/robotics, statistical signal processing, and applications in big data, finance, and marketing.

    Dr. Zhang is a Fellow of the Canadian Academy of Engineering, Fellow of the Engineering Institute of Canada, Fellow of the IEEE, a registered Professional Engineer in Ontario, Canada, and a member of Beta Gamma Sigma Honor Society. He is the general Co-Chair for the IEEE International Conference on Acoustics, Speech, and Signal Processing, 2021. He is the general co-chair for 2017 GlobalSIP Symposium on Signal and Information Processing for Finance and Business, and the general co-chair for 2019 GlobalSIP Symposium on Signal, Information Processing and AI for Finance and Business. He was an elected Member of the ICME steering committee. He is the general chair for ICME2024. He is Editor-in-Chief for the IEEE JOURNAL OF SELECTED TOPICS IN SIGNAL PROCESSING. He is Senior Area Editor for the IEEE TRANSACTIONS ON IMAGE PROCESSING. He served as Senior Area Editor for the IEEE TRANSACTIONS ON SIGNAL PROCESSING and Associate Editor for the IEEE TRANSACTIONS ON IMAGE PROCESSING, the IEEE TRANSACTIONS ON MULTIMEDIA, the IEEE TRANSACTIONS ON CIRCUITS AND SYSTEMS FOR VIDEO TECHNOLOGY, the IEEE TRANSACTIONS ON SIGNAL PROCESSING, and the IEEE SIGNAL PROCESSING LETTERS. He was selected as IEEE Distinguished Lecturer by the IEEE Signal Processing Society and by the IEEE Circuits and Systems Society.

\end{IEEEbiography}

\begin{IEEEbiography}
    [{\includegraphics[width=1in,height=1.25in,clip,keepaspectratio]{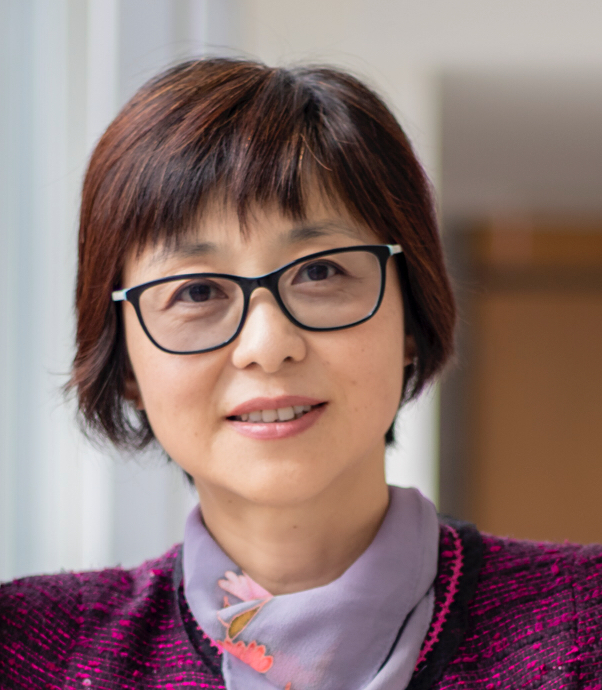}}]{Lian Zhao}
    (Fellow, IEEE) received the Ph.D. degree from the Department of Electrical and Computer Engineering (ELCE), University of Waterloo, Canada, in 2002. She joined the Department of Electrical and Computer Engineering at Toronto Metropolitan University, Canada, in 2003. Her research interests are in the areas of wireless communications, resource management, mobile edge computing, caching and communications, and IoV networks. 
    
    She has been an IEEE Communication Society (ComSoc) and IEEE Vehicular Technology Society (VTS) Distinguished Lecturer (DL); received the Best Land Transportation Paper Award from IEEE Vehicular Technology Society in 2016 and 2024, Best Paper Award from the 2013 International Conference on Wireless Communications and Signal Processing (WCSP) and 2011 ChinaCom, and the Outstanding New Leader Award from IEEE Toronto Section in 2021. 
    
    She has been serving as an Editor for IEEE Transactions on Wireless Communications, IEEE Internet of Things Journal, IEEE Transactions on Vehicular Technology (2013-2021), and an Associate EiC for China Communications since 2023. She served as an Industry Panel co-Chair for Globecom2024,  Wireless Communication Symposium for Globecom 2020 and ICC 2018; Finance co-Chair for 2021 ICASSP; Local Arrangement co-Chair for VTC Fall 2017 and Infocom 2014; co-Chair of Communication Theory Symposium for Globecom 2013. She has been an elected member for the Board of Governor (BoG) of IEEE VTS since 2023. She is a Fellow of IEEE and Asia-Pacific Artificial Intelligence Association, a licensed Professional Engineer in the Province of Ontario.

\end{IEEEbiography}

\end{document}